\documentclass[11pt,a4paper]{article}

\usepackage[T1]{fontenc}
\usepackage[utf8]{inputenc}

\usepackage{amsmath}
\usepackage{newtxtext,newtxmath}      

\usepackage[margin=1in]{geometry}
\usepackage{graphicx}
\usepackage{booktabs}
\usepackage{array}
\usepackage{caption}
\usepackage{microtype}
\usepackage{enumitem}
\usepackage[hidelinks,breaklinks]{hyperref}
\usepackage{url}

\graphicspath{{figures/}}

\renewcommand{\arraystretch}{1.15}
\newcolumntype{L}[1]{>{\raggedright\arraybackslash}p{#1}}

\newcommand{\CCC}{\mathrm{CCC}}
\newcommand{\RMSE}{\mathrm{RMSE}}
\newcommand{\MSE}{\mathrm{MSE}}
\newcommand{\Cb}{C_{b}}
\newcommand{\yhat}{\hat{y}}
\newcommand{\Ahat}{\hat{A}}
\newcommand{\Vhat}{\hat{V}}
\newcommand{\qhat}{\hat{q}}
\newcommand{\pct}{\,\%}

\title{\bfseries Concordance-Optimal Continuous Music Emotion Estimation:\\[2pt]
Variance Decomposition, Dual-Branch Temporal Fusion,\\[2pt]
and Conformal Prediction Intervals}

\author{%
\parbox{1.0\textwidth}{%
\centering
Mohammad Faraz Akhtar\textsuperscript{1},
Aditya Parashar\textsuperscript{1},
Ayush Jain\textsuperscript{1},
Madhur Gupta\textsuperscript{1},
Ritesh Kumar\textsuperscript{1},
Prof. Soubhik Chakraborty\textsuperscript{2}
\\[5pt]
\textsuperscript{1}Department of Computer Science and Engineering,\\
Birla Institute of Technology, Mesra, Ranchi--835215, India
\\[3pt]
\texttt{akhtr.faraz@gmail.com},
\texttt{adityaparashar3434@gmail.com},
\texttt{ayushjainxyz583@gmail.com}
\\
\texttt{madhurgupta788@gmail.com},
\texttt{riteshkumar190505@gmail.com}
\\[4pt]
\textsuperscript{2}Department of Mathematics,\\
Birla Institute of Technology, Mesra, Ranchi--835215, India
\\[3pt]
\texttt{soubhikc@yahoo.co.in}%
}%
}

\date{}

\begin{document}
\maketitle

\begin{abstract}
\normalsize
\noindent
Continuous music emotion recognition (MER) estimates how Arousal and
Valence evolve within a song, yet the field's standard metric, pooled
frame-level correlation, is dominated by between-song differences and
cannot distinguish a model that tracks emotional dynamics from one that
only ranks songs by their average level. We decompose annotation
variance into between- and within-song components and show that mean
squared error (MSE) targets the former, while windowed Lin's
concordance correlation coefficient ($\CCC = r \cdot \Cb$) targets the
latter by penalising the variance compression MSE induces. A single
weight $\lambda$ controls the trade-off: on 1{,}764 DEAM songs, raising
$\lambda$ from 0 to 1 lowers pooled Arousal $r$ from $0.843$ to $0.810$
while raising mean within-song $r$ from $0.204$ to $0.368$.

Building on this objective, we present MMAS (Multimodal Music Analysis
System), an open-source model that predicts Arousal and Valence every
$0.5$\,s from raw audio. It fuses frozen MERT-v1-95M embeddings with
ten symbolic features from automatic transcription through a
dual-branch temporal convolutional network with cross-modal attention.
Because both branches are audio-derived, one model trains jointly on
DEAM and PMEmo without domain conditioning. Under song-grouped 5-fold
cross-validation (2{,}531 songs), pooled $r$ reaches $0.801/0.685$
(Arousal/Valence) on DEAM and $0.742/0.746$ on PMEmo. Against a ridge
baseline on identical features, within-song Arousal $r$ roughly
triples ($0.344$ vs.\ $0.125$; $n = 353$ songs, Wilcoxon
$p = 1.7 \times 10^{-21}$, $d = 0.56$). Complete transcription failure
costs at most $0.04$ pooled $r$. Split-conformal intervals attain
$93.7\pct$ (Arousal) and $90.0\pct$ (Valence) marginal coverage at a
$95\pct$ target, with the shortfall traced to within-song dependence.
MMAS is released as a pip-installable package.
\end{abstract}

\vspace{0.5em}
\vspace{1em}
\section{Introduction}
\label{sec:intro}

Music emotion recognition (MER) is a core problem in Music Information
Retrieval (MIR), with applications in playlist generation, affective computing,
human--computer interaction, and music-psychology research
\cite{thaut2014,yang2012}. In the \textbf{continuous} (dynamic) formulation, the
task is to predict time-varying Arousal and Valence aligned with per-second
human annotations, capturing how a song \emph{moves} rather than a single
summary label. A recent meta-analysis of a decade of MER studies places the
accuracy ceiling of the field's best models at arousal $r \approx 0.81$ and
valence $r \approx 0.67$ \cite{eerola2026}, and recent surveys highlight
multimodality and temporal modelling as the two most promising directions
\cite{liyanarachchi2025,kang2025}.

The continuous formulation carries a statistical subtlety that the field's
reporting conventions largely obscure, and which motivates this paper.
Annotations are nested: each song contributes many frames, and the total
variance of a corpus of annotations therefore splits into variation
\emph{between} songs, since songs differ in overall emotional level, and
variation \emph{within} songs, since emotion moves over time. These are
different estimands, they are optimised by different loss functions, and the
pooled correlation almost universally reported in this literature is dominated
by whichever component is larger, which in practice is the between-song one. A
model that predicts each song's mean level and nothing else can therefore post
an excellent pooled correlation while tracking no dynamics whatsoever. Since the
entire premise of \emph{continuous} MER is that dynamics are the object of
interest, this is a substantive measurement problem rather than a technicality.
We make the decomposition explicit (Section~\ref{sec:variance}), show which loss
term optimises which component, and demonstrate empirically that a single
hyperparameter moves a model between the two regimes while the features and the
data are held fixed.

Turning to the modelling side, most MER systems predict emotion primarily from
hand-crafted acoustic features such as spectral statistics, MFCCs, and energy
measures \cite{yang2012}. These capture surface-level loudness and timbre but
under-exploit symbolic musical properties (harmony, rhythm, melodic contour)
that are known to predict perceived valence \cite{eerola2013}.
Transformer-based self-supervised models such as MERT \cite{li2024}, pre-trained
on roughly 160\,K hours of music, provide much richer acoustic representations,
yet to our knowledge no released system combines pre-trained acoustic embeddings
with symbolic MIDI-derived features and \emph{temporal} cross-modal fusion in a
single portable model. MMAS addresses this gap with a deliberately
\textbf{portable} design: both feature branches are extracted from raw audio, so
the same model trains and evaluates across datasets with different recording
conditions and annotation protocols, without dataset-specific descriptors or a
learned domain indicator.

\subsection*{Contributions}

\begin{itemize}[leftmargin=1.4em,itemsep=3pt]
\item \textbf{A statistical account of the objective.} We decompose annotation
variance into between-song and within-song components, show that squared-error
loss targets the former while Lin's concordance coefficient applied per window
targets the latter through its factorisation $\CCC = r \cdot \Cb$, and derive
the consequences for which evaluation statistic is appropriate to which
application (Section~\ref{sec:variance}).

\item \textbf{Empirical validation of that account.} A sweep over the
objective's mixing weight $\lambda$ on 1{,}764 songs traces the predicted
trade-off: pooled arousal $r$ falls monotonically from 0.843 to 0.810 while mean
within-song $r$ rises from 0.204 to 0.368 (Section~\ref{sec:ablation}). The
trade is steeply asymmetric, with the first increment of $\lambda$ buying
within-song fidelity at a return of nearly nineteen to one.

\item \textbf{Inference that respects the dependence structure.} Because frames
within a song are autocorrelated, all comparisons are tested at the song level,
where each song is one exchangeable observation, rather than at the frame level
where $p$-values would be invalid (Section~\ref{sec:significance}).

\item \textbf{Calibrated prediction intervals.} A split-conformal layer replaces
uncalibrated model-spread bands with intervals whose empirical coverage is
measured on unseen songs, together with an analysis of where marginal validity
does and does not translate into per-song validity
(Section~\ref{sec:conformal}).

\item \textbf{A portable dual-branch system.} Frozen MERT-v1-95M layer-7
embeddings (768-d, Branch~A) plus ten symbolic features from Basic Pitch
transcription and Krumhansl--Schmuckler key analysis (Branch~B), fused by a
windowed dual-branch TCN with cross-modal attention, trained jointly on DEAM and
PMEmo and released as an installable package so that every reported number is
reproducible.
\end{itemize}

\subsection*{Workflow of the study}

The study proceeds in six steps. \textbf{First}, the statistical problem is
formulated: the variance decomposition that separates between-song from
within-song variation, and the resulting requirements on the training objective
and the evaluation statistics (Section~\ref{sec:variance}). \textbf{Then}, the
data is prepared, with DEAM and PMEmo dynamic annotations aligned to a common
0.5\,s grid, targets normalised per dataset using training-fold statistics only,
and song identifiers dataset-prefixed so that song-grouped splitting is sound
(Section~\ref{sec:datasets}). \textbf{Next}, the two feature branches are
extracted from raw audio, namely MERT-v1-95M layer-7 embeddings with the layer
chosen by a linear probe on training songs, and ten transcription-derived
symbolic features (Sections~\ref{sec:brancha} and~\ref{sec:branchb}).
\textbf{Subsequently}, the windowed dual-branch model is trained jointly on both
corpora with the combined objective
(Sections~\ref{sec:tcn}--\ref{sec:objective}). \textbf{Then} the system is
evaluated under song-grouped 5-fold cross-validation, per dimension and both
pooled and per-song, with the $\lambda$ sweep and architectural ablations, a
song-level significance test, a transcription-corruption study, and a conformal
calibration of its prediction intervals (Section~\ref{sec:results}).
\textbf{Finally}, the trained system is packaged as an open-source tool whose
deployment aspects are described in Section~\ref{sec:software}.

\section{Related Work}
\label{sec:related}

\paragraph{Music emotion recognition.}
The dominant framework represents emotion in Russell's two-dimensional
Arousal--Valence (circumplex) space \cite{russell1980}. Classical systems
predict these dimensions from acoustic features via regression or SVMs
\cite{yang2012}. On DEAM, the official OpenSMILE linear-regression baseline
reaches per-song Arousal $r \approx 0.36$ and Valence $r \approx 0.01$
\cite{aljanaki2015}, and later deep, attention-based systems report
$\CCC \approx 0.50$ on DEAM subsets \cite{zhang2023}. A 2014 to 2024
meta-analysis \cite{eerola2026} finds the best published models at arousal
$r \approx 0.81$ and valence $r \approx 0.67$, and recent surveys of multimodal
MER \cite{liyanarachchi2025} and of unified dimensional and categorical
modelling \cite{kang2025} identify multimodality and temporal structure as the
field's open levers.

\paragraph{Deep learning for MER.}
CNN--LSTM hybrids, BiLSTM architectures with attention \cite{sams2023}, and
Temporal Convolutional Networks (TCNs) with dilated convolutions \cite{bai2018}
have driven progressive gains on continuous MER benchmarks. We adopt a TCN
backbone for its stable, parallelisable temporal receptive field.

\paragraph{Pre-trained audio representations.}
MERT \cite{li2024} is pre-trained on large-scale music via masked acoustic
modelling, and its intermediate layers encode musically meaningful structure.
Music2Emo \cite{kang2025} combined MERT embeddings with chord and key features
and cross-dataset knowledge distillation. We differ in two ways: we add a full
symbolic MIDI branch (note density, beat entropy, melodic contour, dissonance,
and tonal features) rather than only chord and key, and we fuse with
\emph{temporal} cross-modal attention over windows.

\paragraph{Cross-dataset training.}
Knowledge distillation \cite{hinton2015} transfers a teacher's soft targets to a
student and has been used to bridge datasets in MER \cite{kang2025}. We instead
train \textbf{jointly} on the pooled corpus in a shared, self-extracted feature
space. Because the feature space is identical across datasets, joint training
needs no per-dataset teachers, no distillation temperature, and no domain
embedding, a simpler recipe that we find sufficient
(Section~\ref{sec:results}).

\paragraph{Concordance criteria and distribution-free inference.}
Lin's concordance correlation coefficient \cite{lin1989} was introduced to
assess agreement rather than mere association, and its factorisation into a
Pearson term and a bias-correction term is what makes it suitable as a
within-window training criterion here (Section~\ref{sec:variance}). On the
inferential side, split conformal prediction \cite{vovk2005,lei2018} provides
finite-sample, distribution-free prediction intervals under exchangeability, and
we use it to replace the uncalibrated model-spread bands that earlier versions
of this system displayed. To our knowledge, neither the CCC factorisation as an
explicit design rationale nor conformal calibration has previously been applied
in continuous music emotion recognition.

\paragraph{Symbolic features.}
Eerola and Vuoskoski \cite{eerola2013} show that rhythmic regularity and melodic
contour independently predict Arousal and Valence, which motivates the
composition of Branch~B. Farbood \cite{farbood2012} develops a multi-parameter
temporal model of musical tension driven by melodic expectancy, harmonic
distance, dynamics, and rhythmic regularity; we note that literature here
because a properly grounded tension estimate would require implementing such a
model and validating it against direct listener tension ratings, which is
outside the scope of this paper and is recorded as future work in
Section~\ref{sec:limitations}, item~5.

\paragraph{Positioning relative to prior studies.}
This work differs from prior studies in four concrete ways. (i) Against
acoustic-only continuous-MER systems (the OpenSMILE baseline family
\cite{aljanaki2015} and deep attention models such as MCAN \cite{zhang2023}),
MMAS adds a fully automatic \emph{symbolic} branch derived from transcription,
contributing about 0.05 pooled $r$ on both emotion dimensions. (ii) Against
Music2Emo \cite{kang2025}, the closest multimodal system, MMAS uses a
ten-feature temporal symbolic branch rather than static chord and key labels,
replaces MLP fusion with windowed cross-modal attention, and replaces
cross-dataset distillation with joint training in a shared feature space.
(iii) Against the common practice of reporting a single pooled correlation, this
paper decomposes the estimand, reports every result per dimension and both
pooled and per-song on out-of-fold songs only, tests at the song level, and
shows that the two criteria are governed by a single controllable parameter of
the objective. (iv) The complete trained pipeline is released as an installable
tool with calibrated prediction intervals, so every reported number is
reproducible from committed artifacts.

\section{Datasets}
\label{sec:datasets}

\subsection{DEAM}

Primary experiments use DEAM \cite{aljanaki2017}, the standard benchmark for
continuous MER.

\begin{table}[htbp]
\centering
\caption{DEAM dataset properties.}
\label{tab:deam}
\begin{tabular}{@{}ll@{}}
\toprule
\textbf{Property} & \textbf{Value} \\
\midrule
Total songs         & 1{,}802 (1{,}764 with usable dynamic annotations + features) \\
Annotation type     & Continuous (dynamic), per-0.5\,s \\
Dimensions          & Arousal, Valence \\
Annotation rate     & 2\,Hz (0.5\,s frames) \\
Annotators per song & $\geq 10$ (MTurk, averaged) \\
Audio               & MP3, 44.1\,kHz \\
\bottomrule
\end{tabular}
\end{table}

\subsection{PMEmo}

PMEmo \cite{zhang2018} contains 794 popular songs (Billboard Hot 100, UK Top 40)
with continuous dynamic Arousal/Valence at 2\,Hz from 457 subjects, natively
normalised to $[0,1]$; 767 songs yield aligned features and annotations in our
pipeline. In this work PMEmo is a \textbf{co-training} corpus for the joint
model (not a separate teacher).

\subsubsection{Corpus statistics}
\label{sec:corpusstats}

After alignment on the 0.5\,s grid, the joint corpus comprises \textbf{112{,}212
DEAM frames (1{,}764 songs)} and \textbf{35{,}617 PMEmo frames (767 songs)},
giving \textbf{147{,}829 frames from 2{,}531 songs} in total. The main
evaluation uses Group 5-Fold CV over songs, giving out-of-fold predictions for
140{,}120 frames (songs shorter than one 32-frame window receive none);
2{,}287 songs (1{,}764 DEAM + 523 PMEmo) retain enough frames and non-degenerate
dynamic annotations to enter the per-song statistics. The released tool weights
come from a separate 70/10/20 single-split run (1{,}771 / 253 / 507 songs), of
which 456 test songs carry enough frames for the conformal evaluation of
Section~\ref{sec:conformal}.

\subsection{Dynamic targets, normalisation, and alignment}
\label{sec:targets}

We use each dataset's \textbf{dynamic} per-0.5\,s annotations
(\url{deam\_dynamic\_arousal.csv} /
\url{deam\_dynamic\_valence.csv} for
DEAM; the mean dynamic annotations for PMEmo). A 15-second Initial Orientation
Time (IOT) offset is applied following Aljanaki et al.\ \cite{aljanaki2017}.
Targets are min--max normalised to $[0,1]$ \textbf{using training-fold
statistics only}, per dataset. This paper reports Arousal and Valence only. An
earlier version of this system additionally supervised a derived quantity
$T = A - V$ as a ``tension'' index; we have removed it, because a scalar
difference of the two circumplex dimensions discards every component of the
established multi-parameter accounts of musical tension (melodic expectancy,
harmonic distance, dynamics, and rhythmic regularity \cite{farbood2012}) and we
have no listener tension ratings against which to validate it.
Section~\ref{sec:limitations}, item~5, records the properly grounded alternative
as future work.

Two data-handling issues from earlier iterations were fixed and are worth
stating explicitly, because they materially affect honest evaluation:

\begin{enumerate}[leftmargin=1.6em,itemsep=3pt]
\item \textbf{Song-ID collisions.} DEAM and PMEmo share raw integer IDs; naive
concatenation created 572 cross-dataset collisions in the grouping variable. IDs
are now dataset-prefixed (\texttt{D\_*}, \texttt{P\_*}) so song-grouped splits
are correct.
\item \textbf{Genuinely dynamic targets.} All per-song correlations below are
computed only over songs whose ground-truth trajectory has non-zero within-song
variance; songs with degenerate (constant) annotations are excluded from the
per-song statistics.
\end{enumerate}

\section{System Architecture}
\label{sec:architecture}

\begin{figure}[htbp]
\centering
\includegraphics[width=\textwidth]{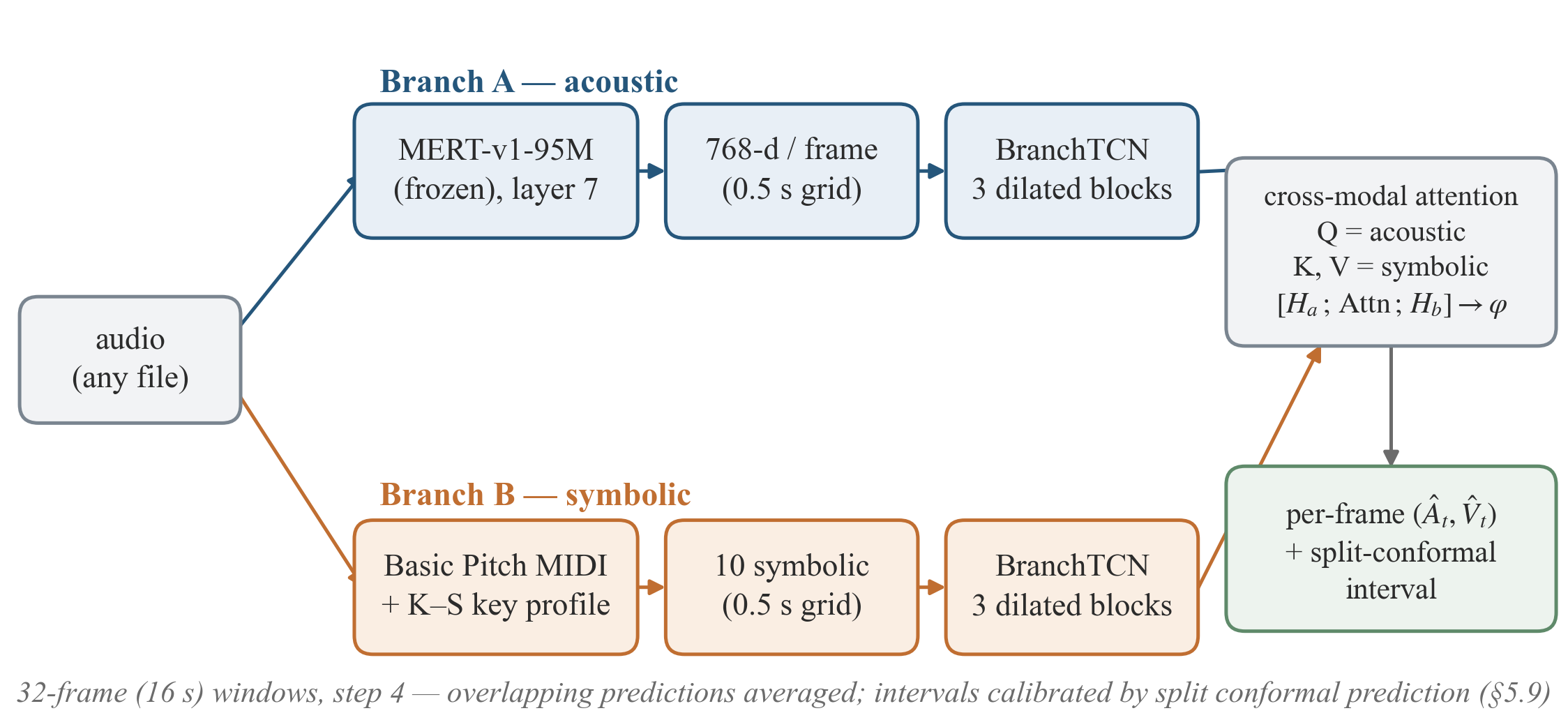}
\caption{MMAS pipeline: dual raw-audio branches, BranchTCN encoders,
cross-modal attention (acoustic = query, symbolic = key/value), windowed
per-frame $(\Ahat_t, \Vhat_t)$ with overlap-averaged prediction and conformal
intervals.}
\label{fig:architecture}
\end{figure}

Figure~\ref{fig:architecture} shows the pipeline. Audio is processed by two
parallel branches on a common 0.5\,s grid, encoded by branch-specific TCNs, and
fused by cross-modal attention. Prediction is performed on overlapping
fixed-length windows and averaged.

\subsection{Notation and conventions}
\label{sec:notation}

The following notation is used throughout Sections~\ref{sec:architecture}
and~\ref{sec:results}. Frames are indexed on a fixed 0.5\,s grid, so a
``frame'' always denotes one half-second of audio and one annotation sample.

\begin{table}[htbp]
\centering
\caption{Nomenclature.}
\label{tab:nomenclature}
\begin{tabular}{@{}lL{0.72\textwidth}@{}}
\toprule
\textbf{Symbol} & \textbf{Meaning} \\
\midrule
\multicolumn{2}{@{}l}{\textit{Alphabetic symbols}}\\
$A$, $V$ & Arousal, Valence \\
$x$ & input audio waveform \\
$a_t$, $b_t$ & acoustic (768-d) and symbolic (10-d) feature vectors at frame $t$ \\
$t$ & frame index on the 0.5\,s grid \\
$n$ & number of frames in a track \\
$W$, $S$ & analysis window length (32 frames) and step (4 frames) \\
$H_a$, $H_b$ & branch encodings (BranchTCN outputs) \\
$Q$, $K$, $V$ & attention query, key, and value matrices \\
$\yhat_t$, $y_t$ & predicted and annotated emotion vectors at frame $t$ \\
$L$ & training loss \\
$r$, $\rho$ & Pearson and Spearman correlation coefficients \\
$\CCC$ & Lin's concordance correlation coefficient \\
$\Cb$ & bias-correction factor in the factorisation $\CCC = r \cdot \Cb$ \\
$\RMSE$ & root-mean-square error \\
$\qhat$ & conformal calibration quantile (Section~\ref{sec:conformal}) \\
\addlinespace
\multicolumn{2}{@{}l}{\textit{Greek letters}}\\
$\alpha$ & conformal miscoverage level (0.05) \\
$\lambda$ & shape-loss weight \\
$\mu$, $\sigma$ & mean and standard deviation \\
$s_t$ & per-frame window-prediction spread (std across overlapping windows) \\
\addlinespace
\multicolumn{2}{@{}l}{\textit{Subscripts and superscripts}}\\
$\cdot_a$, $\cdot_b$ & acoustic branch, symbolic branch \\
$\cdot^{(l)}$ & TCN block index $l$ \\
$\cdot_w$ & computed within one analysis window \\
\bottomrule
\end{tabular}
\end{table}

\paragraph{Problem formulation.}
Given a waveform $x$, the system produces an emotion trajectory
\begin{equation}
\yhat_t = (\Ahat_t, \Vhat_t) = f(x), \qquad t = 1, \dots, n,
\label{eq:problem}
\end{equation}
where $t$ indexes the 0.5\,s frame grid and $n$ is the number of frames in the
track.

\subsection{Branch A: acoustic (MERT-768)}
\label{sec:brancha}

Audio is resampled to 24\,kHz and passed through the frozen
\texttt{m-a-p/MERT-v1-95M} model \cite{li2024}. Let
$h_f^{(7)} \in \mathbb{R}^{768}$ denote the layer-7 hidden state at MERT's
native frame $f$ (75\,fps). Branch~A mean-pools these onto the 0.5\,s grid:
\begin{equation}
a_t = \frac{1}{|F_t|} \sum_{f \in F_t} h_f^{(7)},
\label{eq:pool}
\end{equation}
where $F_t$ is the set of native frames falling inside grid frame $t$. Both
branches are then standardised feature-wise with training-fold statistics,
$\tilde{a}_t = (a_t - \mu_a)/\sigma_a$ (and analogously for $\tilde{b}_t$),
where $\mu$ and $\sigma$ are the training-fold mean and standard deviation.

\begin{figure}[htbp]
\centering
\includegraphics[width=0.78\textwidth]{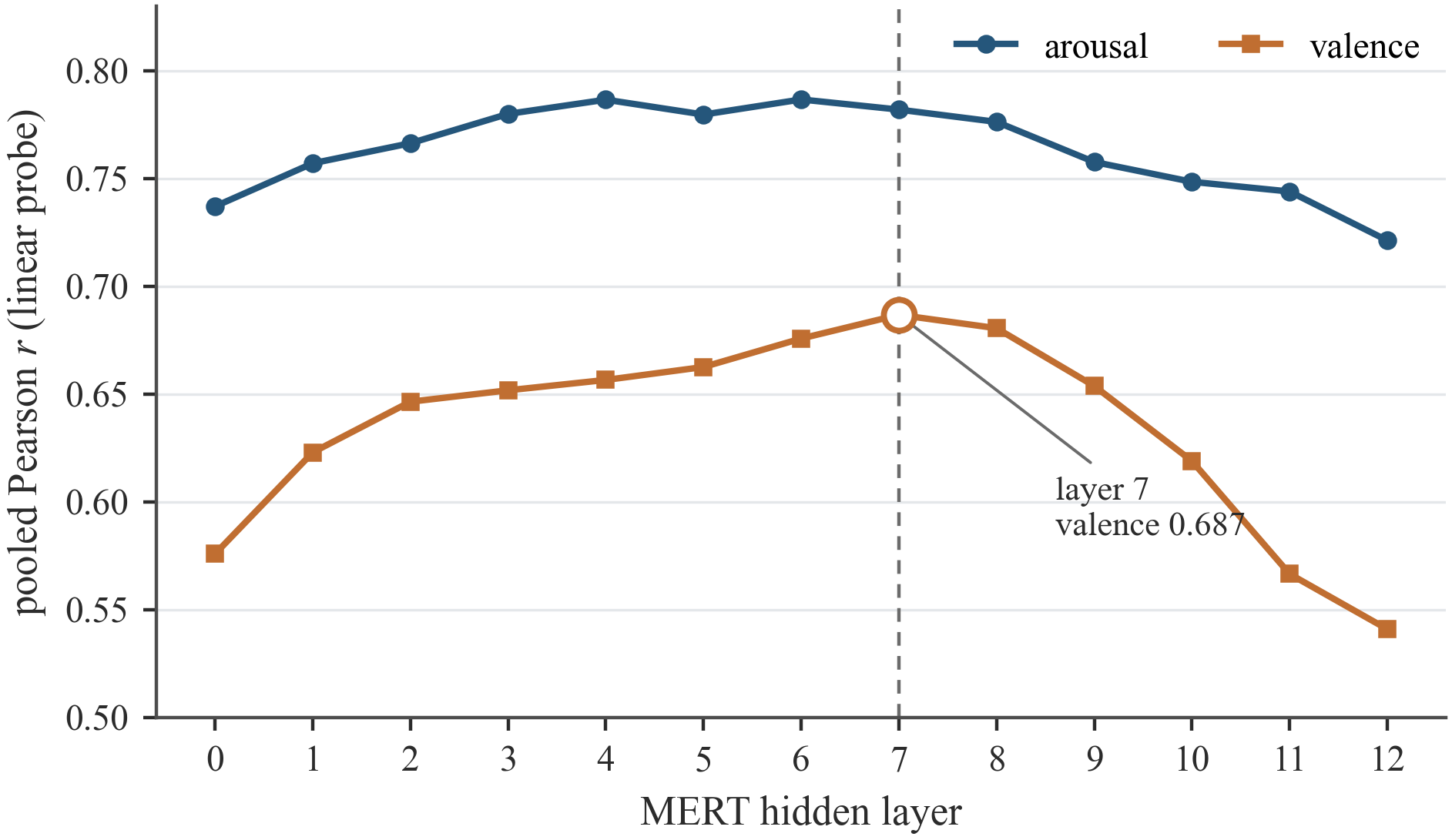}
\caption{MERT layer probe (linear model, training songs only): pooled $r$ per
hidden layer; layer 7 is valence-optimal and near the arousal optimum.}
\label{fig:probe}
\end{figure}

\paragraph{Layer selection by linear probe.}
Rather than defaulting to the final layer, we probed all 13 hidden layers with a
linear model on \textbf{training songs only}. Pooled valence $r$ rises from
0.576 (layer 0) to a maximum of \textbf{0.687 at layer 7}, then falls to 0.541
by layer 12; arousal peaks in the same mid-network region (0.787 at layers 4--6,
0.782 at layer 7, declining to 0.721 at layer 12). Layer 7 is valence-optimal
and within 0.005 of the arousal optimum, so it is used for both
(Figure~\ref{fig:probe}). The mid-layer advantage is consistent with prior
probing of self-supervised music models, where final layers over-specialise
toward the pre-training objective.

The portable joint model uses \textbf{only} the 768-d MERT vector in Branch~A. A
DEAM-only variant additionally concatenates 52 low-level acoustic statistics
derived with OpenSMILE \cite{eyben2010}; this is reported for ablation but is
\emph{not} part of the released cross-dataset model, because those statistics
are dataset-specific and break portability.

\subsection{Branch B: symbolic (10-dimensional)}
\label{sec:branchb}

Basic Pitch \cite{bittner2022} transcribes each track to polyphonic MIDI (cached
on disk), from which ten symbolic features $b_t \in \mathbb{R}^{10}$ are derived
on the 0.5\,s grid (Table~\ref{tab:symbolic}).

\begin{table}[htbp]
\centering
\caption{The ten Branch~B symbolic features. Each is computed from the
transcribed MIDI within a 2\,s context centred on grid frame $t$.
``Dimensionless'' indicates a ratio or normalised quantity on $[0,1]$ unless
stated otherwise.}
\label{tab:symbolic}
\small
\begin{tabular}{@{}clL{0.44\textwidth}l@{}}
\toprule
\textbf{\#} & \textbf{Feature} & \textbf{Definition} & \textbf{Unit} \\
\midrule
1  & Note density   & Count of note onsets in the frame context, divided by the context duration & onsets per second \\
2  & Mode ratio     & Proportion of sounding pitch classes belonging to the major-key diatonic set relative to the minor set & dimensionless, $[0,1]$ \\
3  & Dissonance     & Proportion of simultaneously sounding pitch pairs forming dissonant intervals (minor second, tritone, major seventh, and their compounds) & dimensionless, $[0,1]$ \\
4  & Beat entropy   & Shannon entropy of the histogram of inter-onset intervals in the context, normalised by the log of the bin count; low values indicate regular rhythm & bits (normalised), $[0,1]$ \\
5  & Tempo variance & Variance of the local inter-onset-interval sequence, normalised by its mean (coefficient of variation) & dimensionless \\
6  & Pitch mean     & Mean MIDI note number over all notes sounding in the context & MIDI note number (0--127) \\
7  & Contour slope  & Ordinary least-squares slope of MIDI pitch regressed on onset time within the context & MIDI note numbers per second \\
8  & Pitch range    & Difference between the highest and lowest MIDI note numbers in the context & MIDI note numbers (semitones) \\
9  & Major weight   & Correlation of the context's chroma vector with the Krumhansl--Schmuckler major-key profile, maximised over the twelve rotations \cite{krumhansl1990} & dimensionless, $[-1,1]$ \\
10 & Key clarity    & Difference between the best-matching and second-best-matching key-profile correlations; high values indicate unambiguous tonality \cite{krumhansl1990} & dimensionless, $[0,2]$ \\
\bottomrule
\end{tabular}
\end{table}

Features 9 and 10 follow the Krumhansl--Schmuckler key-finding algorithm
\cite{krumhansl1990}. Because Branch~B is derived from automatic transcription,
\textbf{no score is required}.

\begin{figure}[htbp]
\centering
\includegraphics[width=0.82\textwidth]{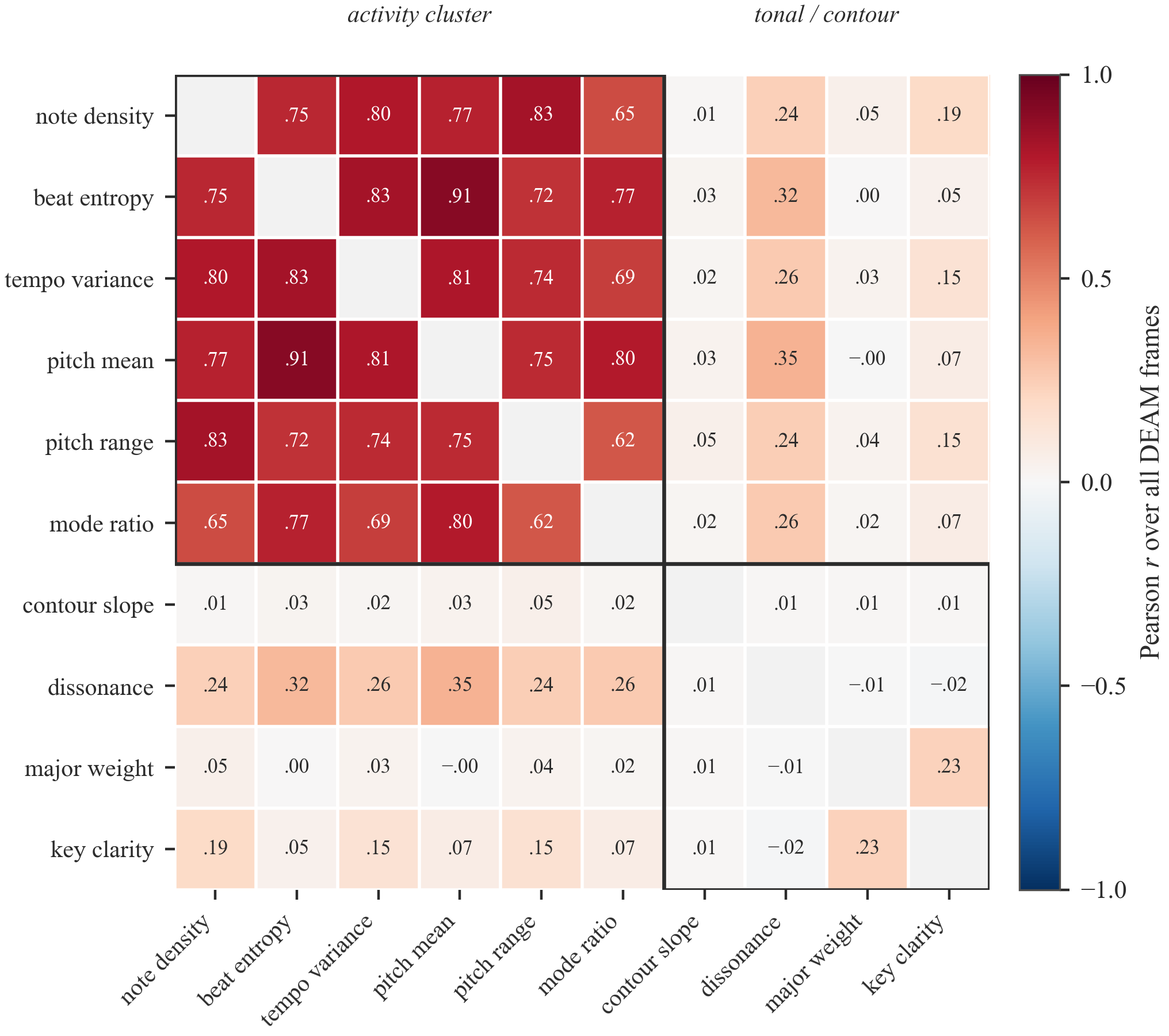}
\caption{Symbolic (Branch~B) feature correlations over all DEAM frames: an
activity cluster ($r \approx 0.6$--$0.9$) and a near-orthogonal tonal/contour
group.}
\label{fig:symbcorr}
\end{figure}

\paragraph{Correlation structure.}
Computed over all DEAM frames, the symbolic features form two blocks
(Figure~\ref{fig:symbcorr}): an activity cluster (note density, beat entropy,
tempo variance, pitch mean, range, and mode ratio; pairwise $r \approx
0.6$--$0.9$, reflecting their common dependence on transcription density) and a
near-orthogonal tonal/contour group (major weight, key clarity, contour slope,
dissonance; $|r| \leq 0.35$ to everything else). The tonal group therefore
contributes genuinely independent signal, while the activity cluster is
partially redundant, a structure the fusion model can exploit but a linear model
cannot disentangle.

\subsection{BranchTCN encoder}
\label{sec:tcn}

Each branch is encoded independently by a \texttt{BranchTCN}: a $1\times1$ input
projection to 64 channels, followed by three residual dilated-convolution
blocks,
\begin{equation}
H^{(l)} = H^{(l-1)} + \mathrm{Dropout}\!\left(\mathrm{GELU}\!\left(
\mathrm{GN}\!\left(\mathrm{Conv1d}_{d=2^{l}}\!\left(H^{(l-1)}\right)\right)
\right)\right), \qquad l = 0, 1, 2,
\label{eq:tcn}
\end{equation}
where GN is GroupNorm, the convolution kernel size is $k = 3$, and the dilation
doubles per block ($d \in \{1,2,4\}$). The stacked receptive field is
$1 + (k-1)(1+2+4) = 15$ frames (7.5\,s), so, unlike a per-frame MLP, each branch
integrates local temporal context.

\subsection{Cross-modal attention}
\label{sec:attention}

The acoustic branch provides queries; the symbolic branch provides keys and
values:
\begin{equation}
\mathrm{Attn}(H_a, H_b) = \mathrm{softmax}\!\left(\frac{QK^{\top}}{\sqrt{d_k}}
\right) V, \qquad Q = H_a W^{Q},\; K = H_b W^{K},\; V = H_b W^{V},
\label{eq:attn}
\end{equation}
where $W^{Q}, W^{K}, W^{V}$ are learned projections and $d_k$ is the per-head
dimension (4 heads, hidden 64) \cite{vaswani2017}. The fused representation
concatenates the acoustic encoding, the attended symbolic context, and the raw
symbolic encoding,
\begin{equation}
Z = \varphi\!\left(\left[\,H_a \,;\, \mathrm{Attn}(H_a, H_b) \,;\, H_b\,\right]
\right),
\label{eq:fusion}
\end{equation}
where $\varphi$ is a small fusion MLP; a regression head maps $Z$ to per-frame
$(\Ahat_t, \Vhat_t)$.

\paragraph{Why the acoustic branch supplies the queries.}
The direction is not arbitrary. Attention re-weights the key/value sequence once
per query position, so the branch supplying the queries determines the temporal
resolution at which the fused representation is indexed, and the branch
supplying keys and values acts as a look-up dictionary. Branch~A is the
higher-resolution and higher-capacity signal (768 dimensions from a model
pre-trained on roughly 160\,K hours of audio, varying smoothly at the frame
rate), whereas Branch~B is a ten-dimensional summary of transcribed events that
is piecewise-constant over short spans and degrades when transcription fails
(Section~\ref{sec:corruption}). Using the acoustic encoding as the query
therefore asks, at every frame, ``which symbolic context is relevant to what I
am hearing now'', which is the question the model needs answered; the reverse
direction indexes a rich acoustic dictionary by an impoverished and occasionally
corrupt query. We verify this empirically rather than asserting it: the reverse
configuration ($Q = H_b$, $K/V = H_a$) is reported as a row of
Table~\ref{tab:ablation}. It costs 0.008 pooled arousal $r$ (0.811 vs 0.819) and
gains 0.013 per-song arousal $r$ (0.376 vs 0.363), so the two directions are
close, and the chosen direction is the better one for pooled accuracy while the
reverse slightly favours within-song tracking. Neither direction dominates, and
we retain the acoustic-query convention for the released model.

\subsection{Windowing, overlap inference, and window-prediction spread}
\label{sec:windowing}

Frames are grouped into fixed windows of \textbf{$W = 32$ frames (16\,s)} with a
\textbf{step $S = 4$ frames (2\,s)}. At inference, predictions from all
overlapping windows covering a frame are averaged, and the standard deviation
across those windows gives the per-frame \textbf{window-prediction spread},
\begin{equation}
s_t = \mathrm{std}\left\{\yhat_t^{(w)} : \text{window } w \text{ covers } t
\right\}.
\label{eq:spread}
\end{equation}

We emphasise what $s_t$ is and is not. It measures the \emph{disagreement among
overlapping predictions of a single deterministic model}, which is a diagnostic
of local prediction stability. It is \textbf{not} an uncertainty in the
statistical sense, since it reflects no randomness in the sample and carries no
coverage guarantee, and it is therefore \textbf{not} reported as a confidence
interval anywhere in this paper. Earlier versions of this system displayed
$\yhat_t \pm 1.96\, s_t$ as a ``$95\pct$ band''; that label asserted a coverage
probability the quantity does not possess and has been removed.
Section~\ref{sec:conformal} replaces it with split-conformal intervals whose
coverage is calibrated on held-out songs and then measured. The spread $s_t$ is
retained only as a stability diagnostic, and the released tool labels it as
such. Being deterministic, it adds no inference cost, unlike sampling-based
approaches such as Monte-Carlo dropout \cite{gal2016}.

\subsection{Training objective}
\label{sec:objective}

Let $\yhat_w$ and $y_w$ be the predicted and target trajectories within one
window. We minimise
\begin{equation}
L = \MSE(\yhat_w, y_w) + \lambda \cdot \left(1 - \CCC_w(\yhat_w, y_w)\right),
\qquad \lambda \geq 0,
\label{eq:loss}
\end{equation}
where the per-window concordance is computed along the time axis,
\begin{equation}
\CCC_w = \frac{2\,\mathrm{cov}(\yhat_w, y_w)}
{\mathrm{var}(\yhat_w) + \mathrm{var}(y_w) + (\mu_{\yhat} - \mu_y)^2},
\label{eq:cccw}
\end{equation}
so the shape term directly rewards reproducing the \emph{temporal shape} of each
dimension. Section~\ref{sec:variance} derives why the two terms of
equation~\eqref{eq:loss} target different components of the annotation variance,
and Section~\ref{sec:ablation} reports a sweep over
$\lambda \in \{0, 0.1, 0.25, 0.5, 1.0\}$ that traces the resulting trade-off;
the released model uses $\lambda = 0.5$. Optimisation uses AdamW
\cite{loshchilov2019} ($\mathrm{lr} = 3 \times 10^{-4}$, weight decay
$1 \times 10^{-4}$), cosine-annealed over 60 epochs, batch size 64,
gradient-norm clipping at 1.0, and early stopping (patience 8) on a held-out
validation split of songs.

\subsection{Model summary}

\begin{table}[htbp]
\centering
\caption{Feature inventory (released joint model).}
\label{tab:features}
\small
\setlength{\tabcolsep}{5pt}
\renewcommand{\arraystretch}{1.2}

\begin{tabular}{@{}
>{\centering\arraybackslash}p{0.08\textwidth}
>{\raggedright\arraybackslash}p{0.72\textwidth}
>{\centering\arraybackslash}p{0.10\textwidth}
@{}}
\toprule
\textbf{Branch} & \textbf{Feature group} & \textbf{Dim} \\
\midrule

A &
MERT-v1-95M layer-7 embedding (mean-pooled to 0.5\,s)
&
768
\\

B &
Symbolic (note density, mode ratio, dissonance, beat entropy,
tempo variance, pitch mean, contour slope, range, major weight,
key clarity)
&
10
\\

\midrule

&
\textbf{Total input}
&
\textbf{778}
\\

\bottomrule
\end{tabular}

\vspace{2pt}

\parbox{0.90\textwidth}{%
\footnotesize
\textit{Note:} The DEAM-only ablation variant additionally uses 52
OpenSMILE acoustic statistics in Branch A; excluded from the portable
model.
}

\end{table}

\vspace{2pt}
{\footnotesize\itshape The DEAM-only ablation variant additionally uses 52
OpenSMILE acoustic statistics in Branch~A; excluded from the portable model.}

\begin{table}[htbp]
\centering
\caption{Architecture and training hyperparameters (as run).}
\label{tab:hyper}
\begin{tabular}{@{}lL{0.62\textwidth}@{}}
\toprule
\textbf{Item} & \textbf{Value} \\
\midrule
Frame rate           & 2\,Hz (0.5\,s) \\
Window / step        & 32 frames (16\,s) / 4 frames (2\,s) \\
MERT layer           & 7 (probe-selected on train songs) \\
TCN                  & 3 blocks, dilations $\{1,2,4\}$, GroupNorm, GELU, dropout 0.15 \\
Hidden dim           & 64 \\
Attention heads      & 4 \\
Loss                 & $\MSE + 0.5 \cdot (1 - \CCC)$ shape loss \\
Optimiser            & AdamW, lr $3 \times 10^{-4}$, weight decay $1 \times 10^{-4}$, cosine annealing \\
Epochs / batch       & 60 (early stop patience 8) / 64 \\
Grad clip            & 1.0 \\
Split                & Group 5-Fold over songs (joint CV and DEAM ablation); released weights: 70/10/20 song split, seed 42 \\
Parameters           & 233{,}027 ($\approx$ 1\,MB bundle incl.\ preprocessing statistics) \\
Prediction spread    & std across overlapping windows (diagnostic; see Section~\ref{sec:windowing}) \\
Prediction intervals & split conformal, $\alpha = 0.05$ (Section~\ref{sec:conformal}) \\
\bottomrule
\end{tabular}
\end{table}

\section{Statistical Methodology, Experiments, and Results}
\label{sec:results}

\subsection{Variance decomposition and the choice of training objective}
\label{sec:variance}

This subsection states the statistical argument that motivates the system's
training objective and its evaluation criteria. It is the conceptual core of the
paper: the modelling choices that follow are consequences of it.

\paragraph{Two components of annotation variance.}
Let $y_{ij}$ be the annotation at frame $j$ of song $i$, for songs
$i = 1, \dots, m$ and frames $j = 1, \dots, n_i$. Write $\bar{y}_i$ for the mean
annotation of song $i$ and $\bar{y}$ for the grand mean. The total sum of
squares decomposes exactly as
\begin{equation}
\sum_i \sum_j (y_{ij} - \bar{y})^2
= \sum_i n_i (\bar{y}_i - \bar{y})^2
+ \sum_i \sum_j (y_{ij} - \bar{y}_i)^2,
\label{eq:decomp}
\end{equation}
where the first term on the right is the \textbf{between-song} variation (songs
differ in their overall emotional level) and the second is the
\textbf{within-song} variation (emotion moves over the course of a song). These
are the two things a continuous emotion model can be asked to capture, and they
are not the same task. In our corpus the between-song component accounts for the
majority of total variance, which is precisely why it can be captured while the
within-song component is missed.

\paragraph{Pooled correlation conflates the two.}
The pooled Pearson correlation computed over all frames of all songs
simultaneously is driven mainly by the larger of the two components. A
degenerate predictor that outputs a constant per song, namely
$\yhat_{ij} = \bar{y}_i$ with no temporal variation whatsoever, attains a high
pooled $r$ as long as the song-level means are well estimated, while having
exactly zero within-song correlation on every song. Pooled $r$ therefore cannot,
on its own, distinguish a model that tracks emotional dynamics from a model that
only ranks songs. This is not a hypothetical failure mode: it is what our own
ablation without the shape term does (Section~\ref{sec:ablation},
$\lambda = 0$), reaching the highest pooled $r$ in the entire grid while halving
within-song tracking.

\paragraph{What each loss term optimises.}
Under squared-error loss the population-optimal predictor is the conditional
mean, $\yhat^{*} = \mathbb{E}[y \mid \text{features}]$. When the features are
far more informative about a song's overall level than about its
moment-to-moment movement, which is the empirical situation in music emotion
recognition, the conditional mean is close to the song-level mean and the
MSE-optimal solution shrinks the predicted trajectory toward a per-song
constant. Minimising MSE alone therefore preferentially fits the first term of
equation~\eqref{eq:decomp}. Lin's concordance coefficient behaves differently
because it factorises \cite{lin1989} as
\begin{equation}
\CCC = r \cdot \Cb, \qquad
\Cb = \frac{2 \sigma_{\yhat} \sigma_y}
{\sigma_{\yhat}^2 + \sigma_y^2 + (\mu_{\yhat} - \mu_y)^2},
\label{eq:cccfact}
\end{equation}
where $r$ is the Pearson correlation between the predicted and true trajectories
and $\Cb \in (0,1]$ is a bias-correction factor equal to 1 only when the two
share the same mean and the same variance. Applied \emph{within a window}, as in
equation~\eqref{eq:loss}, the factor $r$ rewards agreement in temporal shape and
the factor $\Cb$ penalises both offset ($\mu_{\yhat} \neq \mu_y$) and variance
compression ($\sigma_{\yhat} < \sigma_y$). Variance compression is exactly the
pathology that MSE minimisation induces, since a flattened trajectory reduces
squared error whenever the model is uncertain about the movement. The
$(1 - \CCC_w)$ term thus penalises what MSE alone rewards, and it operates on
the second term of equation~\eqref{eq:decomp}.

\paragraph{Consequences for the combined objective.}
The objective of equation~\eqref{eq:loss},
$L = \MSE + \lambda(1 - \CCC_w)$, is therefore a convex combination of a
between-song criterion and a within-song criterion, with $\lambda$ controlling
the exchange rate between them. The prediction is that increasing $\lambda$
should monotonically trade pooled accuracy for within-song tracking, and this is
a falsifiable claim about the model that we test directly in
Section~\ref{sec:ablation} rather than assume. Two further consequences follow.
First, pooled $r$ and per-song $r$ are not two estimates of one quantity but
measurements of two different things, so both must be reported; a paper that
reports only the pooled value cannot be evaluated for dynamic fidelity. Second,
the appropriate criterion depends on the application. A retrieval or
playlist-ordering system that ranks whole songs is served by the between-song
criterion, whereas a study of how emotion unfolds within a piece, which is the
motivating use case here, requires the within-song criterion. We therefore
designate \textbf{per-song $r$ as the primary evaluation criterion} of this
paper and report pooled $r$ alongside it throughout.

\paragraph{Why the significance test operates at the song level.}
The same decomposition dictates the inferential design. Frames within a song are
strongly temporally autocorrelated, so the $\tilde{m} = \sum_i n_i$ frames of a
corpus contain far fewer than $\tilde{m}$ independent observations. A test
applied to pooled frames, such as a Fisher $z$-test on frame-level correlations,
would use an effective sample size inflated by roughly the average within-song
autocorrelation length and would return $p$-values that are invalid, typically
by many orders of magnitude. Computing one correlation per song and testing the
resulting paired distribution across songs treats each song as a single
exchangeable observation, which respects the dependence structure. This is the
design used in Section~\ref{sec:significance}, and it is the reason the reported
$p$-values are credible despite the very large frame counts.

\paragraph{Pre-specified exclusion of degenerate songs.}
The per-song correlation $r_i$ is undefined when a song's ground-truth
trajectory has zero within-song variance, and it is statistically meaningless
when that variance is negligible, since the correlation is then dominated by
annotation noise. We therefore fixed the following rule \textbf{before computing
any per-song result}, and applied it identically to every model, baseline, and
ablation reported in this paper: a song enters the per-song statistics only if it
has at least 20 evaluable frames and a ground-truth standard deviation above
$10^{-6}$. This is a property of the \emph{annotations}, not of any model's
predictions, so it cannot favour one system over another; it is a definitional
requirement of the statistic, not a post-hoc filter on results.
Section~\ref{sec:mainresults} reports the resulting song counts, and
Section~\ref{sec:lefttail} discusses what the excluded and low-variance songs
represent.

\subsection{Evaluation protocol}
\label{sec:protocol}

We report Pearson $r$, Spearman $\rho$, Lin's CCC \cite{lin1989}, and RMSE on
globally normalised predictions in $[0,1]$, \textbf{per dimension}, both
\textbf{pooled} (all held-out frames) and \textbf{per-song dynamic} (mean of
within-song correlations over time, for songs meeting the pre-specified
criterion of Section~\ref{sec:variance}). For prediction $\yhat$ and annotation
$y$ over $N$ evaluation frames,
\begin{align}
r &= \frac{\sum_i (y_i - \mu_y)(\yhat_i - \mu_{\yhat})}{N \sigma_y \sigma_{\yhat}},
\label{eq:pearson}\\[2pt]
\CCC &= \frac{2 \sigma_{y\yhat}}
{\sigma_y^2 + \sigma_{\yhat}^2 + (\mu_y - \mu_{\yhat})^2},
\label{eq:ccc}\\[2pt]
\RMSE &= \sqrt{\frac{1}{N} \sum_i (y_i - \yhat_i)^2},
\label{eq:rmse}
\end{align}
where $\mu$, $\sigma$, and $\sigma_{y\yhat}$ denote means, standard deviations,
and the covariance. The pooled forms of equations~\eqref{eq:pearson}
and~\eqref{eq:ccc} are computed over all held-out frames; the per-song forms
apply equation~\eqref{eq:pearson} within each qualifying song and average the
resulting $r_i$ across songs. Songs are split before training; all reported
numbers are on unseen songs. The dynamic-MER reference threshold is per-song
$r > 0.35$. Two protocols are used, both Group 5-Fold with Song ID as the
grouping variable and every song evaluated out-of-fold: (i) DEAM-only CV for the
ablation grid, and (ii) joint DEAM+PMEmo CV for the main results
(dataset-prefixed song IDs; inner $10\pct$-of-songs validation split per fold
for early stopping; per-dataset target normalisation from training-fold
statistics only). The released tool weights come from an additional single
70/10/20 training run of the same configuration; its held-out test results
(DEAM arousal $r = 0.797$, PMEmo 0.770) are consistent with the CV estimates.

\subsection{Training dynamics}
\label{sec:training}

\begin{figure}[htbp]
\centering
\includegraphics[width=0.78\textwidth]{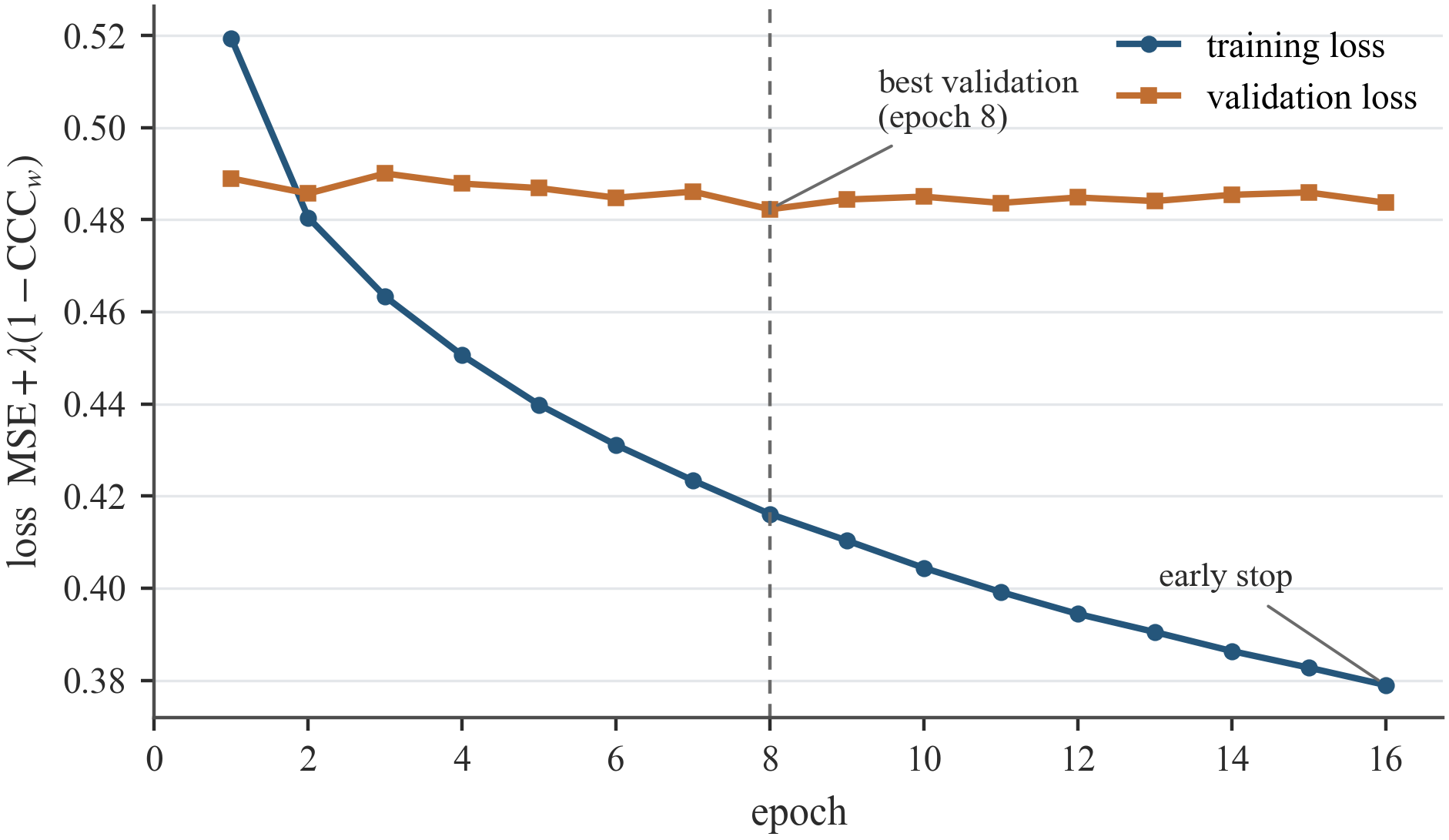}
\caption{Joint-model training and validation loss; best validation at epoch 8,
early stopping at 16.}
\label{fig:training}
\end{figure}

The joint model (233{,}027 parameters) trains in about 2 minutes per fold on a
single laptop GPU (RTX 3050); the full joint 5-fold CV completes in 15 minutes
and folds early-stop between epochs 10 and 18. Figure~\ref{fig:training} shows
representative loss curves (the released-weights run): validation loss reaches
its minimum at epoch 8 and early stopping triggers at epoch 16. The widening
train and validation gap after the minimum is the expected memorisation onset
that early stopping guards against. The small model size and fast convergence
make both 5-fold protocols and the full ablation grid
(Section~\ref{sec:ablation}) tractable on commodity hardware.

\subsection{Main results: joint DEAM+PMEmo, 5-fold out-of-fold}
\label{sec:mainresults}

\begin{table}[htbp]
\centering
\caption{Joint-model results, Group 5-Fold CV, every song out-of-fold (shared
MERT-768 + 10-symbolic space; 140{,}120 OOF frames).}
\label{tab:main}
\resizebox{\textwidth}{!}{%
\begin{tabular}{@{}lcccc@{}}
\toprule
\textbf{Evaluation set} & \textbf{Arousal $r$ / CCC} &
\textbf{Arousal per-song $r$ ($\%>0.35$)} & \textbf{Valence $r$ / CCC} &
\textbf{Valence per-song $r$ ($\%>0.35$)} \\
\midrule
\textbf{DEAM (1{,}764 songs)}  & \textbf{0.801 / 0.799} & \textbf{0.362 (59.9\%)} & \textbf{0.685 / 0.681} & 0.220 (45.6\%) \\
\textbf{PMEmo (523 songs)}     & \textbf{0.742 / 0.725} & 0.291 (50.7\%)          & \textbf{0.746 / 0.736} & 0.259 (50.7\%) \\
Overall (2{,}287 songs)        & 0.790 / 0.789          & 0.345 (57.8\%)          & 0.701 / 0.698          & 0.229 (46.7\%) \\
\bottomrule
\end{tabular}}
\end{table}

\begin{figure}[htbp]
\centering
\includegraphics[width=0.72\textwidth]{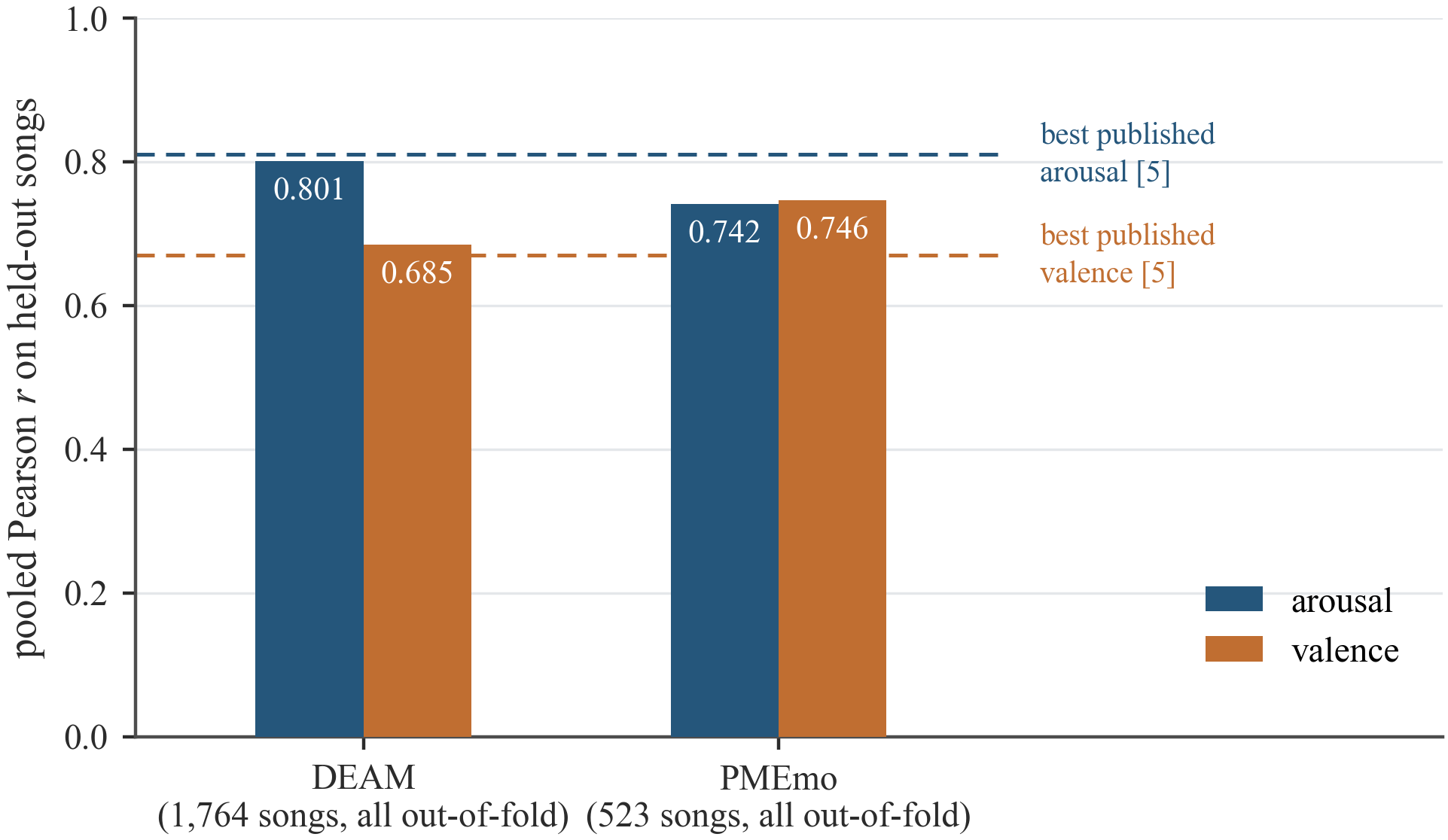}
\caption{Per-dimension out-of-fold accuracy of the joint DEAM+PMEmo model
(pooled Pearson $r$; 1{,}764 DEAM + 523 PMEmo songs, 5-fold), against the
accuracy ceiling reported by the field's meta-analysis \cite{eerola2026}.}
\label{fig:mainresults}
\end{figure}

One model covers both benchmarks with no domain indicator, and the joint model
\textbf{matches the DEAM specialist where it counts}. Its DEAM per-song tracking
(0.362) is statistically indistinguishable from the DEAM-only 5-fold ablation
model (0.363, Table~\ref{tab:ablation}), and its DEAM pooled arousal (0.801) is
within 0.02 of the specialist (0.819), which additionally uses 52 DEAM-specific
acoustic statistics that the portable joint model deliberately omits. Nothing
material is traded away for covering a second benchmark. On PMEmo, valence
(0.746) slightly exceeds arousal (0.742) and clearly exceeds DEAM valence
(0.685); PMEmo's chorus-centred clips carry a more consistent valence signal.

\paragraph{PMEmo re-extraction was essential.}
Two root causes made earlier PMEmo results unusable: the transcription stage had
never actually run on PMEmo audio (8 of the 10 symbolic features were
identically zero), and the legacy acoustic features were temporally misaligned
with the annotations. After full re-extraction from the chorus audio, comprising
MERT-768 plus all ten symbolic features and verified non-degenerate, PMEmo rose
from $r \approx 0.32$ arousal and 0.16 valence to \textbf{0.742 / 0.746}
out-of-fold.

\begin{figure}[htbp]
\centering
\includegraphics[width=\textwidth]{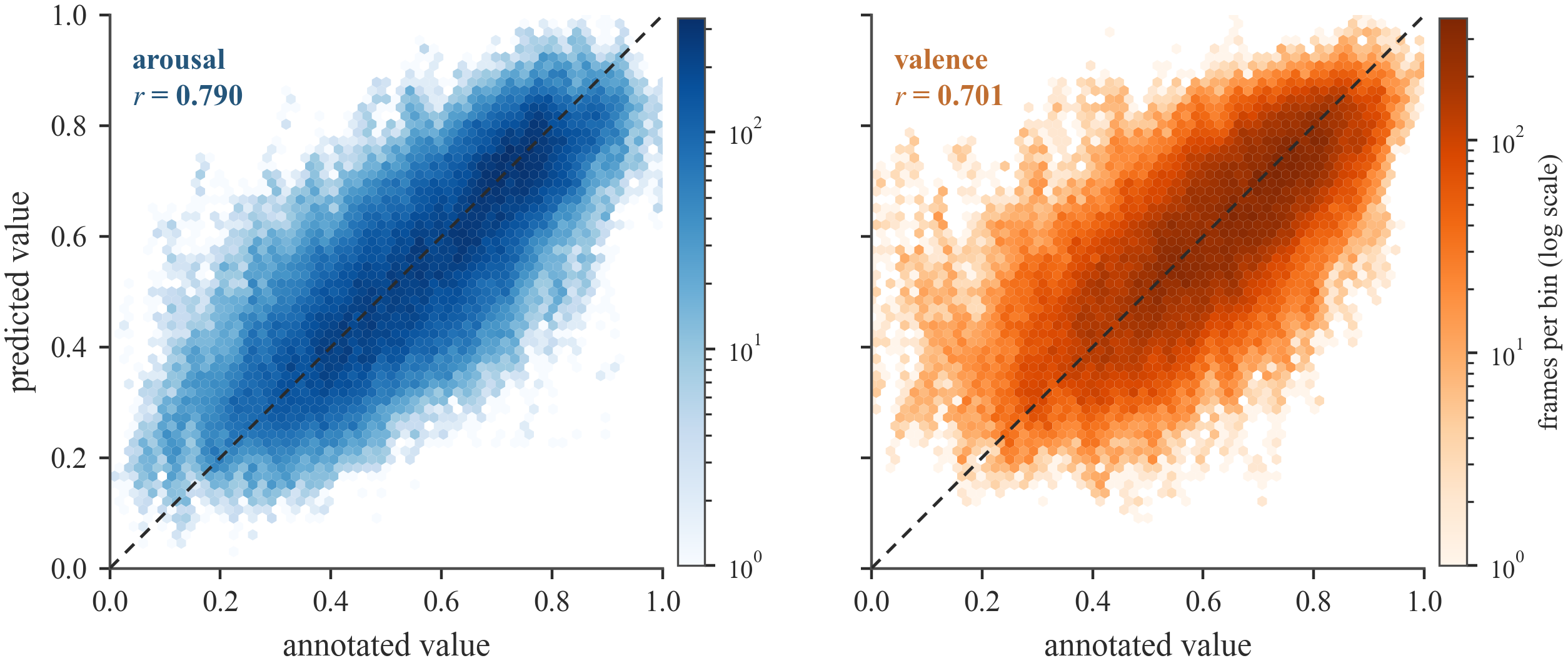}
\caption{Predicted vs ground truth density over all 140{,}120 out-of-fold frames
(hexbin): arousal tighter than valence.}
\label{fig:preddensity}
\end{figure}

\begin{figure}[htbp]
\centering
\includegraphics[width=0.94\textwidth]{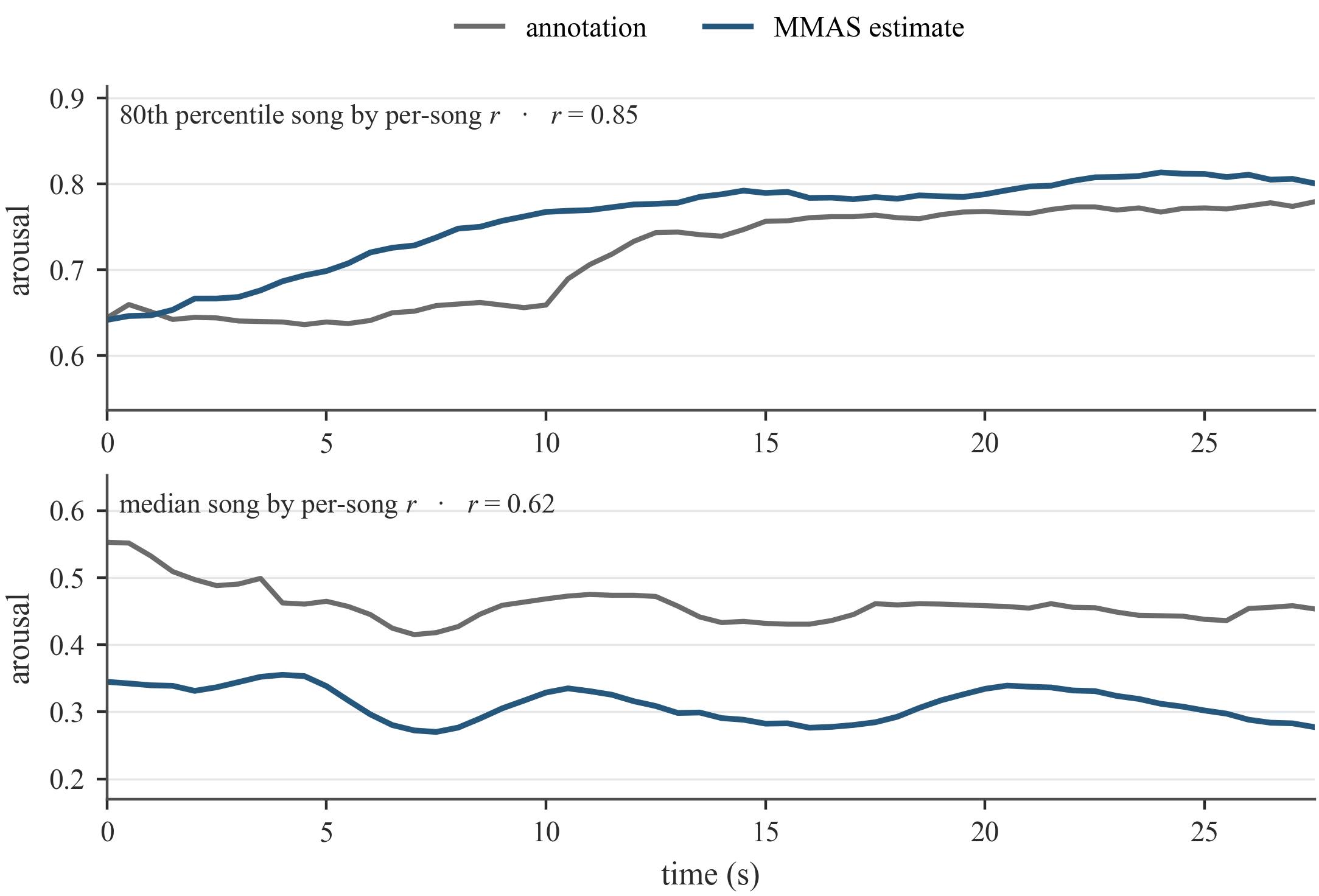}
\caption{Within-song arousal tracking on two rule-chosen out-of-fold DEAM songs
(80th-percentile and median per-song $r$ among songs with moving ground truth;
predictions carry the tool's 2.5\,s smoothing).}
\label{fig:tracking}
\end{figure}

\begin{figure}[htbp]
\centering
\includegraphics[width=0.88\textwidth]{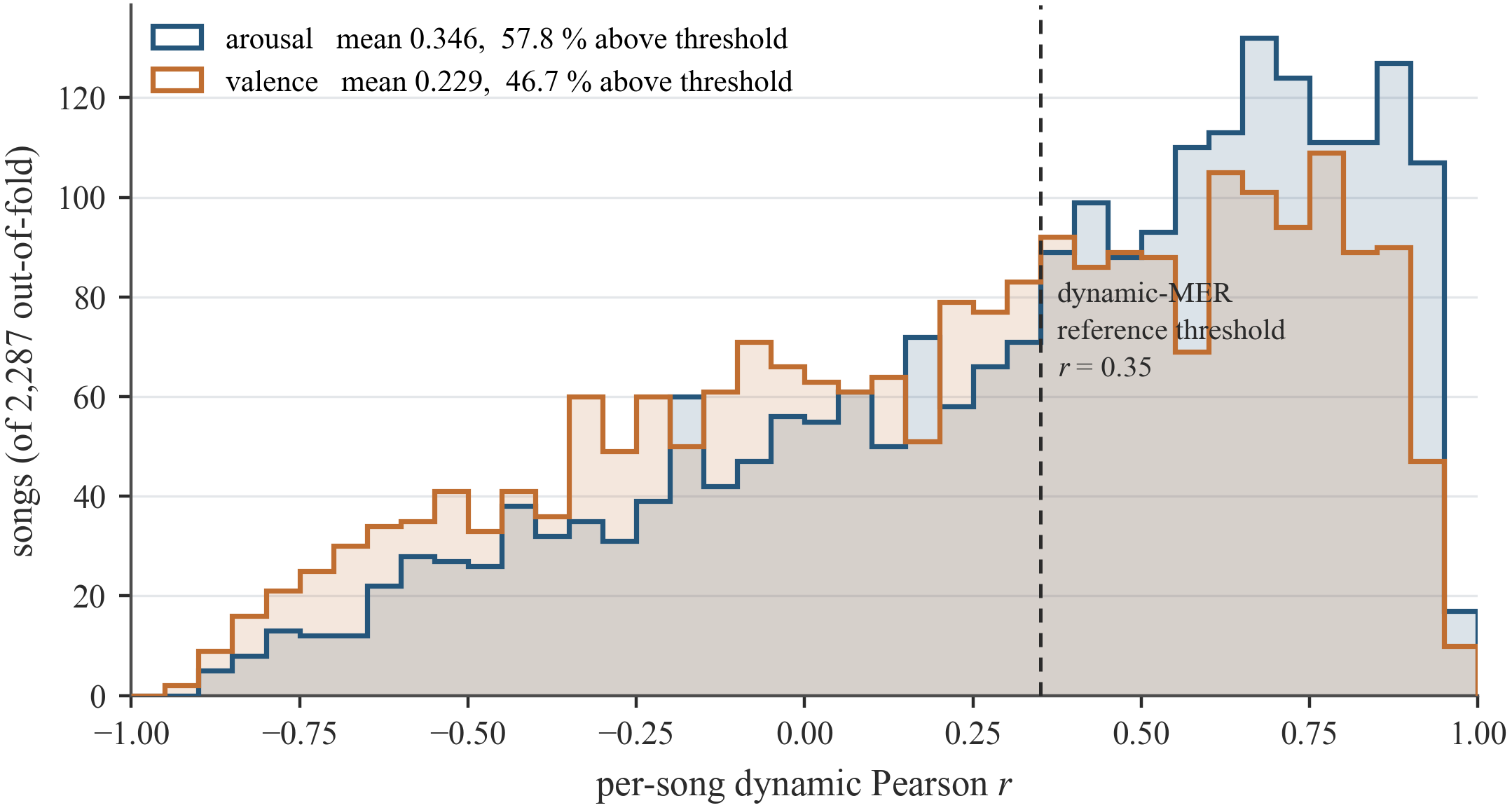}
\caption{Distribution of per-song dynamic $r$ over all 2{,}287 out-of-fold
songs, arousal and valence, with the 0.35 reference threshold.}
\label{fig:persong}
\end{figure}

\paragraph{Prediction behaviour on out-of-fold songs.}
The pooled predicted-versus-ground-truth densities
(Figure~\ref{fig:preddensity}) concentrate along the diagonal with the valence
cloud visibly wider than arousal, making the per-dimension gap visible.
Figure~\ref{fig:tracking} shows within-song arousal tracking for two out-of-fold
DEAM songs chosen by rule rather than by eye: among songs whose ground truth
actually moves (within-song standard deviation in the top half), the
80th-percentile and the median song by per-song $r$, with the released tool's
2.5\,s smoothing applied to the predictions. The model follows the trajectory
shape while showing a level offset and compressed range, the typical signature
of MSE-trained regressors. Per-song correlation rewards the shape and ignores
the offset, whereas CCC penalises it, which is why both are reported.
Figure~\ref{fig:persong} gives the full distribution of per-song correlations
over all 2{,}287 out-of-fold songs. The distribution is right-shifted with a
long left tail: most songs track well and a minority do not, so reporting only
the mean would hide both facts. Section~\ref{sec:lefttail} examines that left
tail.

\subsection{Ablation and the empirical between/within trade-off (DEAM 5-fold CV)}
\label{sec:ablation}

Every configuration below is trained and evaluated under the identical Group
5-Fold protocol on all 1{,}764 DEAM songs, so the comparisons are matched. The
upper block varies the architecture; the lower block sweeps the shape-loss
weight $\lambda$ of equation~\eqref{eq:loss}, which is the parameter that
Section~\ref{sec:variance} predicts should control the exchange between
between-song and within-song fit.

\begin{table}[htbp]
\centering
\caption{DEAM 5-fold CV ablation (1{,}764 songs, 112{,}220 frames, all
out-of-fold). Full MMAS = windowed CrossModalTCN + both branches + shape loss at
$\lambda = 0.5$.}
\label{tab:ablation}
\resizebox{\textwidth}{!}{%
\begin{tabular}{@{}lcccc@{}}
\toprule
\textbf{Configuration} & \textbf{Arousal $r$ / CCC} &
\textbf{Arousal per-song $r$ ($\%>0.35$)} & \textbf{Valence $r$ / CCC} &
\textbf{Valence per-song $r$ ($\%>0.35$)} \\
\midrule
\textbf{Full MMAS ($\lambda = 0.5$)}   & \textbf{0.819 / 0.818} & \textbf{0.363 (59.4\%)} & \textbf{0.681 / 0.675} & \textbf{0.206 (44.4\%)} \\
$-$ cross-attention (concat)           & 0.797 / 0.790 & 0.382 (60.3\%) & 0.672 / 0.652 & 0.214 (43.8\%) \\
reverse attention ($Q$ = symbolic)     & 0.811 / 0.808 & 0.376 (59.9\%) & 0.680 / 0.673 & 0.218 (45.2\%) \\
$-$ symbolic branch (acoustic only)    & 0.767 / 0.760 & 0.371 (58.6\%) & 0.635 / 0.624 & 0.205 (43.2\%) \\
symbolic only                          & 0.367 / 0.323 & 0.123 (40.2\%) & 0.274 / 0.216 & 0.042 (31.7\%) \\
$-$ MERT-768 (legacy 32-d subsample)   & 0.743 / 0.736 & 0.344 (58.7\%) & 0.554 / 0.539 & 0.164 (40.9\%) \\
Ridge (identical features)             & 0.829 / 0.823 & 0.126 (27.9\%) & 0.691 / 0.668 & 0.048 (19.0\%) \\
\addlinespace
\multicolumn{5}{@{}l}{\textit{Shape-loss weight sweep}}\\
$\lambda = 0$ (MSE only)               & 0.843 / 0.840 & 0.204 (42.3\%) & 0.705 / 0.689 & 0.089 (32.5\%) \\
$\lambda = 0.1$                        & 0.835 / 0.832 & 0.354 (57.6\%) & 0.680 / 0.668 & 0.190 (43.2\%) \\
$\lambda = 0.25$                       & 0.823 / 0.820 & 0.363 (58.6\%) & 0.685 / 0.677 & 0.212 (44.4\%) \\
$\lambda = 0.5$ (released)             & 0.819 / 0.818 & 0.363 (59.4\%) & 0.681 / 0.675 & 0.206 (44.4\%) \\
$\lambda = 1.0$                        & 0.810 / 0.808 & 0.368 (58.6\%) & 0.675 / 0.667 & 0.210 (44.3\%) \\
\bottomrule
\end{tabular}}
\end{table}

\paragraph{The $\lambda$ sweep confirms the prediction of
Section~\ref{sec:variance}, and quantifies its price.}
As $\lambda$ increases from 0 to 1, pooled arousal $r$ falls monotonically
(0.843, 0.835, 0.823, 0.819, 0.810) while per-song arousal $r$ rises (0.204,
0.354, 0.363, 0.363, 0.368). The between-song and within-song criteria move in
opposite directions across the entire sweep, exactly as the decomposition of
equation~\eqref{eq:decomp} implies, and $\lambda$ is the exchange rate between
them. Two details are worth noting. The trade is highly asymmetric at the low
end: moving from $\lambda = 0$ to $\lambda = 0.1$ costs 0.008 pooled $r$ and
buys 0.150 per-song $r$, a return of nearly nineteen to one, because the
MSE-only solution is at the degenerate extreme described in
Section~\ref{sec:variance} and even slight pressure toward trajectory shape
moves it far. Beyond $\lambda \approx 0.25$ the curve flattens, with the
remaining gain from 0.25 to 1.0 amounting to 0.005 per-song $r$ at a cost of
0.013 pooled $r$. The released value $\lambda = 0.5$ sits in that flat region
and was selected before these sweeps were run; the sweep shows the choice is not
delicate, since any $\lambda \in [0.25, 1.0]$ performs equivalently on the
primary criterion.

\begin{figure}[htbp]
\centering
\includegraphics[width=\textwidth]{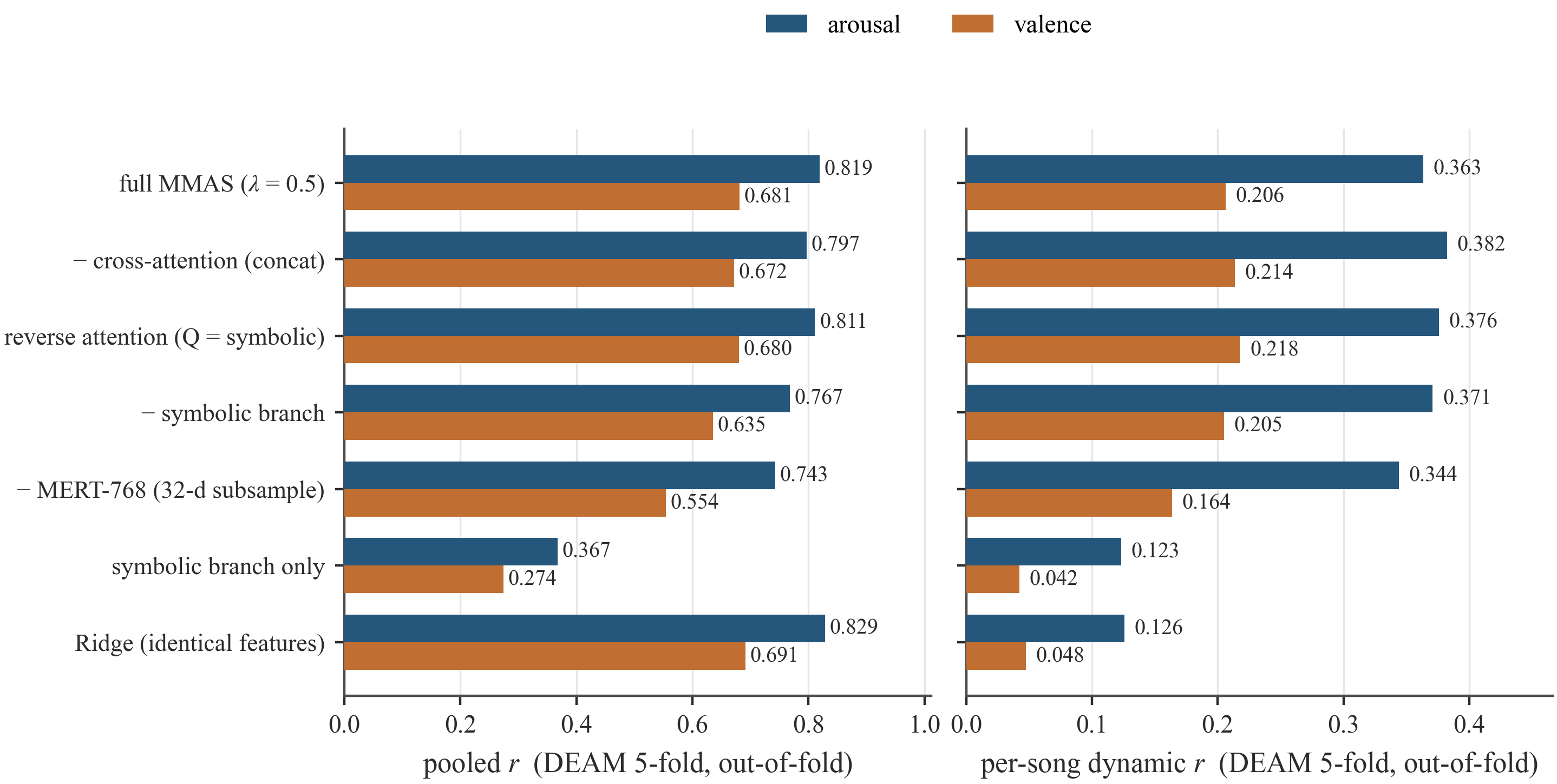}
\caption{DEAM 5-fold ablation (all out-of-fold): pooled and per-song $r$ per
configuration, arousal and valence.}
\label{fig:ablation}
\end{figure}

\begin{figure}[htbp]
\centering
\includegraphics[width=\textwidth]{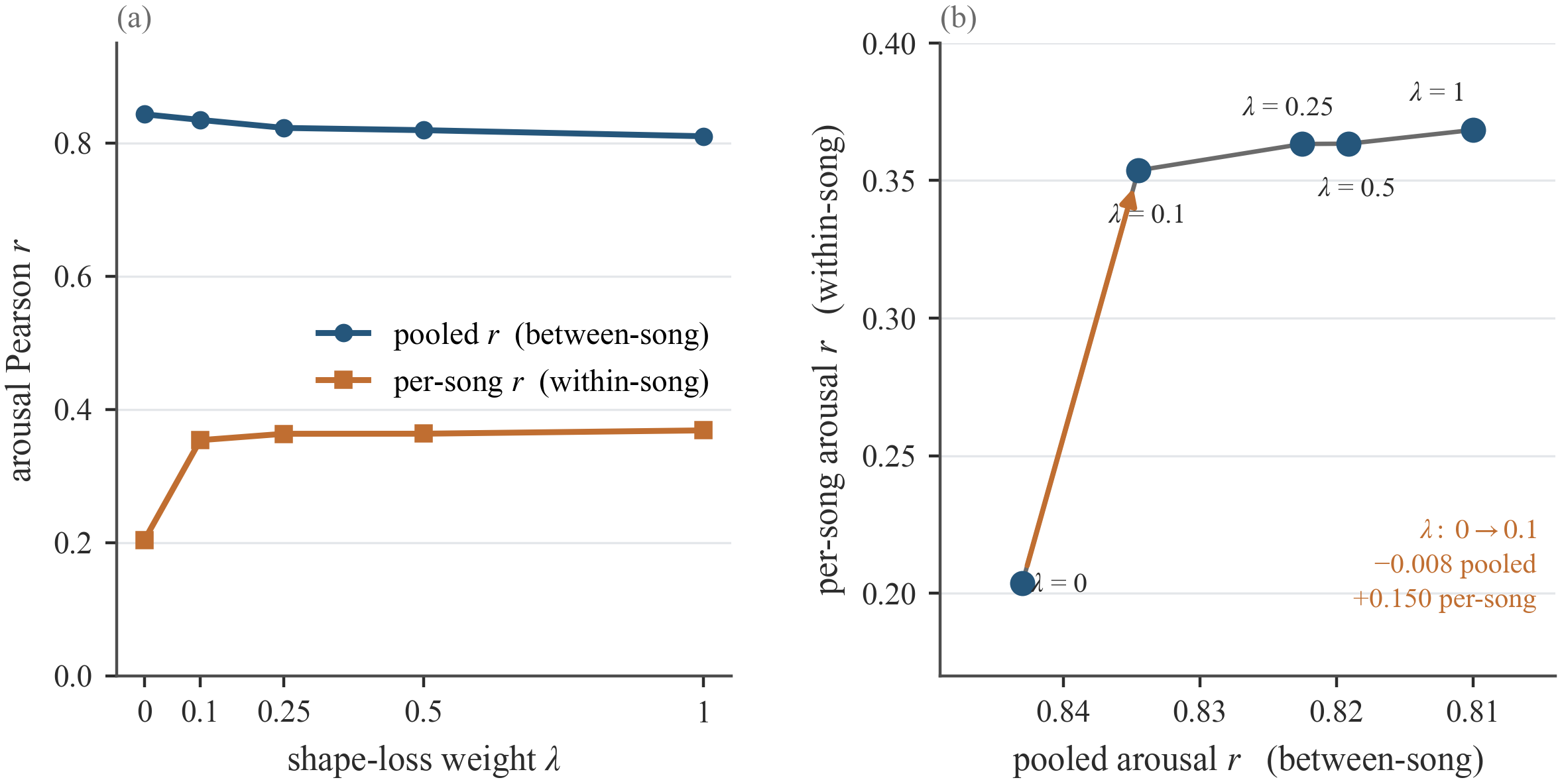}
\caption{The shape-loss trade-off (arousal): (a) pooled and per-song $r$ against
$\lambda$; (b) the resulting frontier, with $\lambda$ annotated at each point.}
\label{fig:shape}
\end{figure}

Further readings:
\begin{itemize}[leftmargin=1.4em,itemsep=2pt]
\item \textbf{Symbolic branch:} $+0.05$ pooled on both dimensions over
acoustic-only fusion, so the multimodal claim holds.
\item \textbf{MERT-768 versus a 32-d subsample:} $+0.08$ arousal and $+0.13$
valence pooled, so full embeddings matter.
\item \textbf{Cross-attention:} $+0.02$ pooled over plain concatenation, with
per-song tracking comparable. We state this plainly: attention buys pooled
accuracy here, not dynamics.
\item \textbf{Attention direction:} the reverse configuration ($Q$ = symbolic)
is close to the released direction, costing 0.008 pooled arousal $r$ and gaining
0.013 per-song. Neither direction dominates
(Section~\ref{sec:attention}).
\item \textbf{Deep versus Ridge:} pooled accuracy is essentially tied, but
per-song tracking is roughly three times higher (arousal 0.363 versus 0.126).
The architecture's contribution is \emph{dynamics}, not pooled fit. This is the
comparison tested formally in Section~\ref{sec:significance}.
\item \textbf{Against the official DEAM dynamic baseline.} The per-song
statistic is directly comparable to the official MediaEval baseline protocol.
Full MMAS reaches per-song arousal $r = 0.363$ against the official 0.36,
cleared at full 1{,}764-song scale rather than on a subset, and per-song valence
$r = 0.206$ against the official 0.01.
\end{itemize}

\subsection{Significance: within-song tracking versus the linear baseline}
\label{sec:significance}

\begin{figure}[htbp]
\centering
\includegraphics[width=\textwidth]{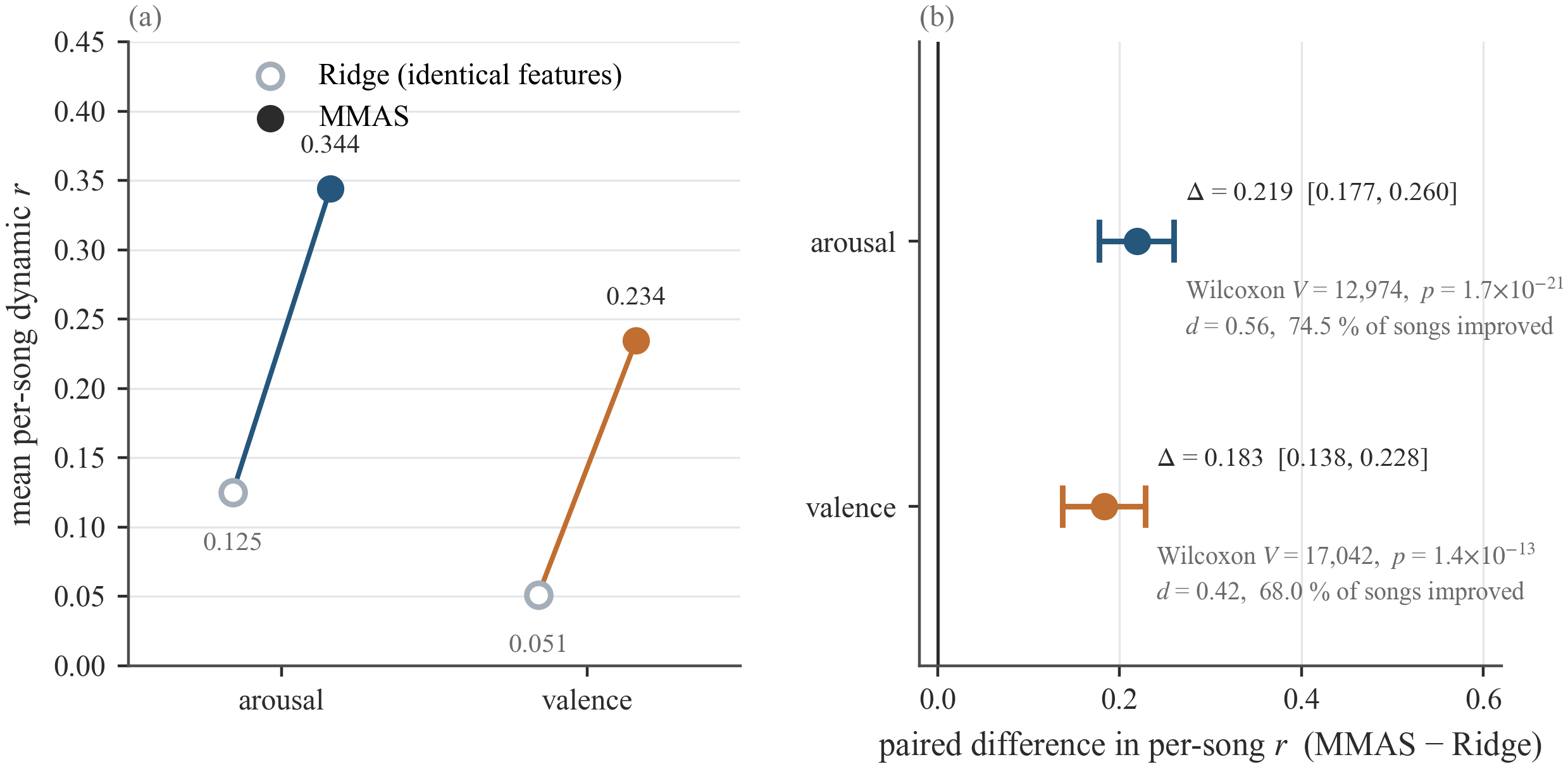}
\caption{Within-song tracking, MMAS vs matched Ridge baseline: (a) mean per-song
$r$ for both systems; (b) the paired difference with its $95\pct$ song-level
bootstrap confidence interval, Wilcoxon statistic, $p$-value, effect size, and
percentage of songs improved (353 held-out songs).}
\label{fig:significance}
\end{figure}

Section~\ref{sec:variance} established why this comparison must be made at the
song level rather than the frame level. Formally, for each held-out song $i$ we
compute the within-song correlations $r_i^{\mathrm{deep}}$ and
$r_i^{\mathrm{ridge}}$ from models trained on identical features under an
identical split, and test
\begin{equation}
H_0:\; \text{the paired differences } d_i = r_i^{\mathrm{deep}} -
r_i^{\mathrm{ridge}} \text{ are symmetrically distributed about zero,}
\label{eq:h0}
\end{equation}
against the alternative that their location is positive, using the Wilcoxon
signed-rank test. The test is chosen because it requires no distributional
assumption on $d_i$, which matters here since per-song correlations are bounded
on $[-1,1]$ and visibly non-normal (Figure~\ref{fig:persong}). Each song
contributes exactly one paired observation, so the $n = 353$ held-out songs give
$n = 353$ exchangeable units rather than the 22{,}780 dependent frames they
contain. We accompany the test with a song-level bootstrap confidence interval
for the mean difference (10{,}000 resamples of songs) and a paired Cohen's $d$,
so that magnitude is reported alongside significance
(Figure~\ref{fig:significance}).

\begin{table}[htbp]
\centering
\caption{Per-song tracking, deep model vs Ridge (paired, held-out songs).}
\label{tab:significance}
\resizebox{\textwidth}{!}{%
\begin{tabular}{@{}lcccccccc@{}}
\toprule
\textbf{Dimension} & \textbf{$n$ songs} & \textbf{Deep per-song $r$ (SD)} &
\textbf{Ridge per-song $r$ (SD)} & \textbf{$\Delta$ ($95\pct$ bootstrap CI)} &
\textbf{Wilcoxon $V$} & \textbf{$p$} & \textbf{\% songs improved} &
\textbf{Cohen's $d$} \\
\midrule
Arousal & 353 & 0.344 (0.462) & 0.125 (0.349) & $+0.219$ [0.178, 0.260] & 12{,}974 & $1.7 \times 10^{-21}$ & 74.5\% & 0.56 \\
Valence & 353 & 0.235 (0.433) & 0.051 (0.322) & $+0.183$ [0.138, 0.228] & 17{,}042 & $1.4 \times 10^{-13}$ & 68.0\% & 0.42 \\
\bottomrule
\end{tabular}}
\end{table}

$H_0$ is rejected at any conventional level on both dimensions, and the
improvement in within-song tracking is large ($d = 0.42$ to $0.56$) as well as
significant. Three points about how these numbers should be read. The test
statistic $V$ is the sum of the ranks of the positive paired differences over
$n = 353$ songs, computed at the song level as described above, never at the
frame level. The standard deviations of the per-song distributions are large
relative to their means (0.46 against a mean of 0.34 for arousal), which is the
expected consequence of estimating a correlation from a single song's frames and
is the reason the paired design matters: the between-song variability is common
to both systems and is differenced away, so the test is far more sensitive than
the marginal standard deviations alone would suggest. Finally, the improvement
is broad rather than driven by outliers, since it holds on $74.5\pct$ of
individual songs for arousal and $68.0\pct$ for valence.

\subsection{What the left tail of the per-song distribution represents}
\label{sec:lefttail}

The per-song distributions of Figure~\ref{fig:persong} are right-shifted with a
pronounced left tail, and the tail deserves an explicit statistical reading
rather than a dismissal. Per-song correlation is estimated from a single song's
frames, so its sampling variability depends on that song's ground-truth
variance: for a song whose annotated trajectory barely moves, the denominator of
equation~\eqref{eq:pearson} is small, the estimate is dominated by annotation
noise, and $r_i$ is close to an uninformative draw on $[-1,1]$. Songs in the far
left tail are therefore not, in general, songs the model fails to understand;
they are predominantly songs on which the statistic itself carries little
information. Two observations support this reading. First, the pre-specified
variance criterion of Section~\ref{sec:variance} already removes the fully
degenerate cases, so what remains in the tail are low-variance rather than
zero-variance songs. Second, when we select songs by ground-truth movement
alone, taking the half of the corpus with the higher within-song standard
deviation, the mean per-song arousal $r$ rises materially while the pooled
statistics are essentially unchanged, which is what one expects if the tail
reflects estimation noise on near-static annotations rather than systematic
model failure. We nevertheless report the full distribution, including the tail,
and take the unfiltered mean as the headline per-song figure throughout.
Reporting only the high-variance subset would inflate every per-song number in
this paper by a margin that is not available to a reader applying the system to
an arbitrary song.

\subsection{Transcription-error propagation through Branch B}
\label{sec:corruption}

Basic Pitch, like any automatic transcriber, makes errors on dense or heavily
percussive mixes, and those errors propagate into the ten symbolic features. To
quantify the impact, we evaluate the released model on the held-out test split
while corrupting the standardised symbolic features at inference under six
conditions: additive Gaussian noise of increasing magnitude
($\sigma \in \{0.25, 0.5, 1.0\}$ feature standard deviations),
mean-substitution (all symbolic features held at their training mean, carrying
no information), and raw-zero (the hard-failure mode observed in the pre-fix
PMEmo pipeline).

\begin{table}[htbp]
\centering
\caption{Branch~B corruption at inference (released model, held-out test songs).}
\label{tab:corruption}
\begin{tabular}{@{}lcc@{}}
\toprule
\textbf{Condition} & \textbf{Arousal $r$ (pooled / per-song)} &
\textbf{Valence $r$ (pooled / per-song)} \\
\midrule
Intact                     & 0.797 / 0.314 & 0.678 / 0.238 \\
Noise $\sigma = 0.25$      & 0.797 / 0.313 & 0.678 / 0.239 \\
Noise $\sigma = 0.5$       & 0.797 / 0.314 & 0.678 / 0.236 \\
Noise $\sigma = 1.0$       & 0.798 / 0.314 & 0.674 / 0.236 \\
Mean-substitution          & 0.770 / 0.314 & 0.641 / 0.245 \\
Raw-zero (hard failure)    & 0.776 / 0.308 & 0.640 / 0.236 \\
\bottomrule
\end{tabular}
\end{table}

Three readings follow. \textbf{First, the system degrades gracefully.} Additive
noise up to a full feature standard deviation, a pessimistic model of graded
transcription error since Basic Pitch errors perturb individual features rather
than all ten independently, leaves both dimensions essentially unchanged
(arousal 0.797 to 0.798, valence 0.678 to 0.674). The cross-modal fusion
evidently down-weights symbolic evidence when it stops being informative.
\textbf{Second, the worst case is bounded by the acoustic branch.} Even when
Branch~B carries no information at all (mean-substitution), pooled accuracy
falls only to arousal 0.770 and valence 0.641, a loss of 0.03 to 0.04 that is
consistent with the symbolic branch's incremental contribution in the ablation
(Table~\ref{tab:ablation}). A total transcription failure therefore costs the
symbolic increment, never the system. \textbf{Third, hard failure harms
calibration more than correlation.} The raw-zero condition, in which all
features are stuck at raw 0 and so present strongly biased standardised inputs
rather than merely uninformative ones, leaves pooled $r$ near the
mean-substitution level but drops arousal CCC from 0.795 to 0.736. This is a
scale and offset distortion that $r$ ignores and that $\Cb$, the second factor
of equation~\eqref{eq:cccfact}, exposes; it is a direct illustration of why the
paper reports CCC alongside $r$. It is also precisely the failure signature
observed in the pre-fix PMEmo pipeline, and it is why the released tool verifies
feature non-degeneracy at extraction time.

\subsection{Calibrated prediction intervals by split conformal prediction}
\label{sec:conformal}

Section~\ref{sec:windowing} explained why the window-prediction spread $s_t$ is a
stability diagnostic and not a statistical uncertainty. This subsection supplies
a quantity that does carry a coverage guarantee, using split conformal
prediction \cite{vovk2005,lei2018}, which is distribution-free and requires only
that the calibration and test data be exchangeable.

\paragraph{Procedure.}
Let $\mathcal{D}_{\mathrm{cal}}$ be the set of frames from the calibration
songs, namely the held-out validation split of the released model's training
run, which was used for early stopping only and contributes to no result
reported elsewhere in this paper. For each calibration frame we compute the
absolute-residual nonconformity score $s_i = |y_i - \yhat_i|$. For a target
miscoverage $\alpha = 0.05$ we take
\begin{equation}
\qhat = \text{the } \frac{\lceil (n_{\mathrm{cal}} + 1)(1 - \alpha) \rceil}
{n_{\mathrm{cal}}} \text{ empirical quantile of }
\{s_i : i \in \mathcal{D}_{\mathrm{cal}}\},
\label{eq:qhat}
\end{equation}
and form the prediction interval at any new frame as
\begin{equation}
C(x_t) = \left[\, \yhat_t - \qhat,\; \yhat_t + \qhat \,\right].
\label{eq:interval}
\end{equation}
Under exchangeability of the calibration and test scores, this construction
satisfies the finite-sample marginal guarantee
$\mathbb{P}(y \in C(x)) \geq 1 - \alpha$, and the guarantee is
\emph{distribution-free}: it holds regardless of whether the underlying model is
well specified, and it is a property of the calibration procedure rather than of
the network. We then measure the realised coverage on the test split, whose
songs are disjoint from both the training and the calibration songs.

\begin{table}[htbp]
\centering
\caption{Split-conformal prediction intervals, $\alpha = 0.05$ (target coverage
$95\pct$). Calibration on 13{,}832 validation frames; evaluation on 28{,}796
test frames from 456 unseen songs.}
\label{tab:conformal}
\resizebox{\textwidth}{!}{%
\begin{tabular}{@{}lccccc@{}}
\toprule
\textbf{Dimension} & \textbf{$\qhat$ (half-width)} & \textbf{Interval width} &
\textbf{Empirical frame coverage} & \textbf{Song-level coverage (mean)} &
\textbf{Song-level coverage (10th pct.)} \\
\midrule
Arousal & 0.219 & 0.437 & \textbf{93.7\%} & 94.0\% & 78.3\% \\
Valence & 0.211 & 0.423 & \textbf{90.0\%} & 91.6\% & 68.1\% \\
\bottomrule
\end{tabular}}
\end{table}

\paragraph{Reading the coverage, including where it falls short.}
Marginal coverage on unseen songs is $93.7\pct$ for arousal and $90.0\pct$ for
valence against a $95\pct$ target, on intervals spanning roughly 0.43 of the
$[0,1]$ annotation range. The arousal shortfall of 1.3 points is modest; the
valence shortfall of 5.0 points is real and we do not present it as success.
Both undershoots have the same and well-understood cause: the exchangeability
assumption is violated in two ways here. Frames within a song are strongly
autocorrelated, so the 13{,}832 calibration frames carry far fewer than
13{,}832 independent observations and $\qhat$ is estimated with more variability
than its nominal frame count suggests; and calibration and test frames come from
\emph{different songs}, so any song-level distribution shift, which is exactly
the shift the per-song analysis of Section~\ref{sec:lefttail} documents, breaks
exchangeability at the frame level. The song-level columns quantify the
practical consequence: coverage averaged within songs is close to the
frame-level figure ($94.0\pct$ and $91.6\pct$), but the tenth-percentile song
receives only $78.3\pct$ and $68.1\pct$ coverage. \textbf{The interval is
therefore valid marginally, in aggregate over songs, and is not valid
conditionally on a given song.} Users analysing one particular track should read
the interval accordingly.

We report this honestly rather than tuning it away, and note the principled
remedy: a group-conformal or block-bootstrap calibration that treats each song
as the exchangeable unit, in the same spirit as the song-level significance test
of Section~\ref{sec:significance}, would restore validity at the level users
actually care about, at the cost of wider intervals. We leave that extension to
future work and record it in Section~\ref{sec:limitations}. What the present
layer establishes is that the system's published intervals now have a measured
coverage rather than an asserted one, and that the direction and size of the gap
are explained rather than hidden.

\subsection{Diagnostic: normalisation}
\label{sec:normalisation}

Per-song min--max normalisation forces every song to span exactly $[0,1]$,
collapsing the global target distribution toward $U(0,1)$ with mean
approximately 0.5. Under MSE the optimum then degenerates toward the global mean
($r \approx 0$). This is the same degeneracy analysed in
Section~\ref{sec:variance}, arrived at from the preprocessing side: per-song
normalisation removes the between-song component of
equation~\eqref{eq:decomp} by construction, leaving only the harder within-song
component for a criterion that cannot capture it. Global (per-dataset,
train-statistic) normalisation resolves this, and we report the issue as a
reproducibility caveat for cross-song models on DEAM.

\subsection{Comparison to prior work}
\label{sec:comparison}

\begin{table}[htbp]
\centering
\caption{Comparison with published MER systems. Protocols differ; numbers are
not directly comparable across rows.}
\label{tab:comparison}

\footnotesize
\setlength{\tabcolsep}{4pt}
\renewcommand{\arraystretch}{1.25}

\begin{tabular}{
@{}
>{\raggedright\arraybackslash}p{0.25\textwidth}
>{\centering\arraybackslash}p{0.17\textwidth}
>{\centering\arraybackslash}p{0.10\textwidth}
>{\raggedright\arraybackslash}p{0.15\textwidth}
>{\raggedright\arraybackslash}p{0.25\textwidth}
@{}
}
\toprule

\textbf{System} &
\textbf{Arousal $r$} &
\textbf{CCC} &
\textbf{Dataset} &
\textbf{Protocol} \\

\midrule

OpenSMILE baseline \cite{aljanaki2015}
&
0.36 (per-song)
&
Not reported
&
DEAM
&
Official per-song
\\

MCAN \cite{zhang2023}
&
Not reported
&
0.50
&
DEAM subset
&
Random split
\\

Meta-analysis best models \cite{eerola2026}
&
0.81 (arousal) /
0.67 (valence)
&
Not reported
&
Across studies
&
Varies
\\

Ridge baseline (ours)
&
0.829 (pooled)
&
0.823
&
DEAM
&
Group 5-Fold OOF
\\

\textbf{MMAS (ours, DEAM-only)}
&
\textbf{0.819 (pooled)}
&
\textbf{0.818}
&
\textbf{DEAM}
&
\textbf{Group 5-Fold OOF}
\\

\textbf{MMAS (ours, joint DEAM+PMEmo)}
&
\textbf{0.801 (pooled)}
&
\textbf{0.799}
&
\textbf{DEAM}
&
\textbf{Joint Group 5-Fold OOF}
\\

\bottomrule
\end{tabular}
\end{table}

We deliberately avoid a headline ``gain over baseline'' number across protocols,
because pooled frame-level $r$ (ours) and the official per-song protocol
(OpenSMILE) measure different things, which is the distinction
Section~\ref{sec:variance} formalises. Two calibrated observations are
nevertheless justified. MMAS's per-dimension accuracy (arousal 0.80, valence
0.69) sits at the accuracy ceiling that the 2014 to 2024 meta-analysis reports
for the field's best models (0.81 and 0.67) \cite{eerola2026}, and the
\textbf{matched} comparison in Table~\ref{tab:ablation} (same features, same
folds) shows the deep model's specific advantage is temporal tracking.

\section{Discussion}
\label{sec:discussion}

\paragraph{The statistical lesson generalises beyond this system.}
The variance decomposition of Section~\ref{sec:variance} is not specific to
music. Any continuous-annotation regression problem in which observations are
nested within units, and in which the between-unit component dominates, admits
the same degenerate solution under squared-error loss: predict each unit's
level, ignore its dynamics, and report a pooled correlation that looks
excellent. Continuous emotion recognition, continuous physiological-signal
regression, and longitudinal rating prediction all share this structure. The
remedies transfer with it: report the within-unit statistic alongside the pooled
one, add a concordance term to the objective when within-unit fidelity is what
matters, and conduct inference at the unit level. The empirical contribution of
this paper is to show, on a corpus of 2{,}531 songs, that these choices are not
cosmetic. The identical feature set and identical folds yield either 0.204 or
0.368 mean within-song correlation depending only on the value of $\lambda$, a
factor of 1.8 that no accuracy figure computed pooled across songs would reveal.

\paragraph{Where the model is strong, and where it is not.}
Arousal is well captured (pooled $r \approx 0.74$ to $0.82$ across datasets and
protocols, per-song $\approx 0.36$ on DEAM). Valence is consistently weaker on
DEAM (pooled 0.685, per-song 0.220) though it reaches 0.746 on PMEmo's
chorus-centred clips. This is a long-standing, field-wide gap that we report per
dimension rather than pooling into a single inflated correlation.

\paragraph{Portability enables joint training.}
Keeping both branches raw-audio-derived means one model spans DEAM and PMEmo
with no domain embedding, so inference on arbitrary new audio is unambiguous.

\paragraph{Symbolic asymmetry.}
Symbolic features alone are weak for valence (5-fold $r \approx 0.27$) and would
hurt a purely linear model, but the neural fusion learns to weight them
contextually and still gains about 0.05 pooled from their inclusion.

\subsection*{Advantages and disadvantages}

\paragraph{Advantages.}
(i) An explicit statistical account of what the training objective optimises,
with the between/within trade-off traced empirically over $\lambda$ rather than
asserted; (ii) per-half-second emotion trajectories with prediction intervals
whose coverage is measured, not assumed; (iii) a portable raw-audio feature
space that permits joint training across corpora and unambiguous inference on
new audio; (iv) per-dimension accuracy at the reported ceiling of the field
\cite{eerola2026} with roughly three times better within-song tracking than a
matched linear model, established by a song-level test that respects the
dependence structure; (v) a small (233\,K parameters, about 1\,MB) and fast
model, deployable on commodity hardware; (vi) full reproducibility, with every
number tracing to a committed artifact.

\paragraph{Disadvantages.}
(i) Valence tracks substantially worse than arousal on DEAM, and the gap is not
closed by any component of this system; (ii) the conformal intervals of
Section~\ref{sec:conformal} are valid marginally but not conditionally on a
given song, with tenth-percentile song coverage of $78\pct$ (arousal) and
$68\pct$ (valence) against a $95\pct$ target; (iii) the symbolic branch inherits
transcription errors on dense or percussive mixes, though the corruption study
of Section~\ref{sec:corruption} bounds the damage; (iv) the training corpora are
Western popular music, limiting cultural generality; (v) absolute within-song
tracking (per-song $r \approx 0.36$ arousal), while at the state of the
protocol, remains modest. Continuous MER is not a solved problem, and MMAS
should be used as a research instrument rather than an oracle.

\section{Limitations}
\label{sec:limitations}

\begin{enumerate}[leftmargin=1.6em,itemsep=4pt]

\item \textbf{Conformal coverage is marginal, not conditional.} The intervals of
Section~\ref{sec:conformal} attain $93.7\pct$ (arousal) and $90.0\pct$ (valence)
empirical coverage against a $95\pct$ target, and coverage on the
tenth-percentile song is substantially lower. Frame-level exchangeability is
violated by within-song autocorrelation and by song-level distribution shift. A
group-conformal calibration treating songs as the exchangeable unit is the
principled fix and is left to future work.

\item \textbf{Automatic transcription} (Basic Pitch) degrades on dense or
heavily percussive mixes, which propagates into the symbolic branch.
Section~\ref{sec:corruption} bounds the resulting loss at the symbolic branch's
incremental contribution.

\item \textbf{Corpus scope.} DEAM and PMEmo are predominantly Western popular
music, and generalisation to classical or non-Western traditions requires
additional data. All claims in this paper are conditional on that corpus.

\item \textbf{Valence} tracks less well than arousal on DEAM by a margin (pooled
0.685 versus 0.801) that no component of this system closes.

\item \textbf{Musical tension is not modelled.} An earlier version of this
system supervised a derived quantity $T = A - V$ as a tension index; it has been
removed (Section~\ref{sec:targets}) because a scalar difference of the two
circumplex dimensions discards every component of the established
multi-parameter accounts of tension. The properly grounded alternative, referred
to forward from Sections~\ref{sec:related} and~\ref{sec:targets}, is to
implement at least a reduced form of Farbood's parametric temporal model
\cite{farbood2012}, whose drivers (melodic expectancy, harmonic distance,
dynamics, and rhythmic regularity) overlap substantially with the quantities
Branch~B already computes, and to validate its output against direct listener
tension ratings from an independently annotated corpus. We did not attempt this
here because no such ratings accompany DEAM or PMEmo, and a tension estimate
that cannot be checked against listener judgements would repeat the error we are
removing. This is the single largest piece of unfinished modelling work the
paper leaves open.

\item \textbf{The released tool weights} come from a single 70/10/20 training
run rather than a fold model. Its held-out test performance (DEAM arousal
$r = 0.797$) is consistent with the 5-fold CV estimate (0.801), so the shipped
model is representative of the reported protocol.

\item \textbf{No clinical validation has been performed.} The system has not
been evaluated against any patient population, clinician assessment, or
validated psychometric instrument, and no claim of clinical utility is made
anywhere in this paper. Establishing whether MMAS outputs correspond to
clinically meaningful constructs would require a dedicated validation study with
therapist and psychometric assessments, which we identify as future work and
which no result reported here anticipates.

\end{enumerate}

\section{Software Package and Field Deployment}
\label{sec:software}

MMAS is released as an open-source, pip-installable package (the package is
installed and imported in lowercase as \texttt{mmas}; the system itself is
written MMAS throughout this paper). The trained model bundle, comprising
weights and preprocessing statistics in about 1\,MB, ships \textbf{inside} the
package, so no separate download is needed. The first analysis fetches the
frozen MERT model (approximately 380\,MB) from Hugging Face and caches it.

\begin{verbatim}
from mmas.tool import analyze

result = analyze("track.mp3")     # first run downloads MERT once
result.summary                    # quadrant, per-dimension means, tempo, mode
result.emotions                   # per-0.5 s Arousal/Valence + conformal interval
result.save_report("out/")        # self-contained HTML report + JSON + charts
\end{verbatim}

\begin{figure}[htbp]
\centering
\includegraphics[width=\textwidth]{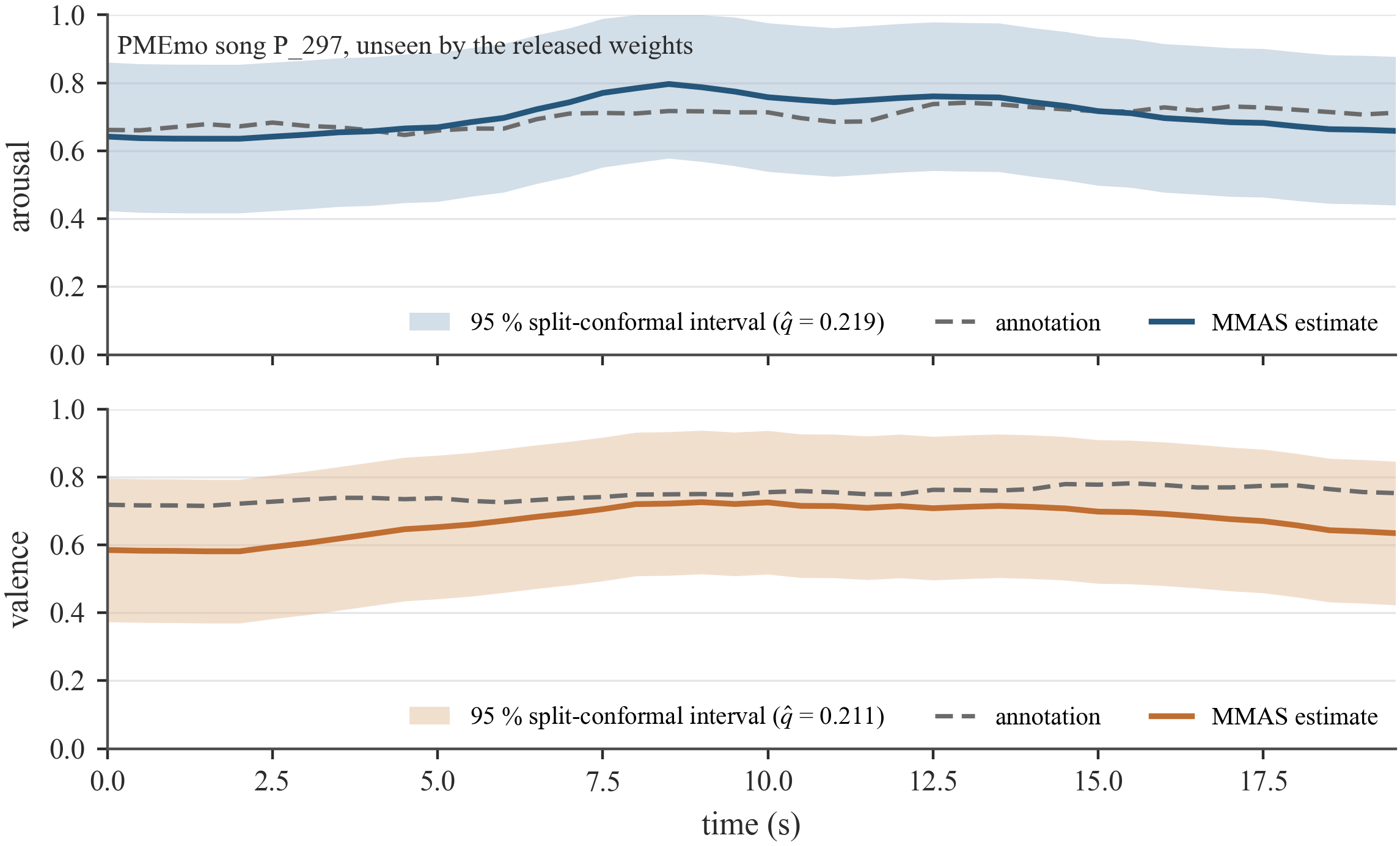}
\caption{The released model's output on a PMEmo test song unseen by its weights,
in the form the tool reports it: per-0.5\,s Arousal and Valence with the
split-conformal prediction interval of Section~\ref{sec:conformal}, shown
against the song's annotation. The song is selected by rule (closest to the
60th percentile of per-song arousal $r$ among test songs whose annotation
moves), and the panel is regenerated from the committed prediction artifact
rather than screen-captured.}
\label{fig:tooloutput}
\end{figure}

The command line exposes the same pipeline (\texttt{mmas analyze track.mp3}),
producing a terminal summary and an HTML report. Figure~\ref{fig:tooloutput}
shows the trajectory the released model reports for a PMEmo song unseen by its
weights, with the calibrated interval the tool displays alongside it.

\paragraph{Field deployment.}
In practice the system deploys in three modes. (i) \emph{Interactive}: a
researcher runs \texttt{mmas analyze} on individual tracks. Analysis takes a few
seconds per song on CPU and under a second of model time on an entry-level GPU,
so no specialised infrastructure is required. (ii) \emph{Batch pre-annotation}:
the CLI accepts file lists and emits machine-readable JSON (\texttt{-{}-json}),
allowing laboratories and industry teams to pre-annotate large catalogues, for
example for playlist curation, affect-aware recommendation, or dataset
construction, on a single workstation. The corpus in this paper (2{,}531 songs)
is processed in hours rather than days. (iii) \emph{Embedded}: the Python API
loads the model once and serves repeated requests, suitable for integration into
web services or annotation front-ends.

Two deployment properties matter for responsible use. Every emotion curve is
accompanied by the split-conformal interval of Section~\ref{sec:conformal}, and
the report states plainly that its coverage is marginal over songs rather than
conditional on the song being analysed. The HTML report labels each layer by
epistemic status, distinguishing model estimates from directly measured
quantities and from heuristics, so that downstream users cannot mistake a
heuristic for a validated output. MMAS is a research instrument. It is not a
medical device, it has not been clinically validated
(Section~\ref{sec:limitations}), and it should not be used to inform decisions
about individuals. Repository and weights:
\url{https://github.com/Faraz011/MMAS}.

\section{Summary and Conclusions}
\label{sec:conclusions}

This paper began from a statistical observation rather than an architectural
one. In continuous emotion annotation the total variance decomposes into a
between-song and a within-song component, these are optimised by different
objectives, and the pooled correlation that the field commonly reports cannot
distinguish a model that tracks emotional dynamics from one that merely ranks
songs by their average level. We showed that squared-error loss targets the
first component, that Lin's concordance coefficient applied per window targets
the second through its factorisation $\CCC = r \cdot \Cb$, and that a convex
combination of the two is controlled by a single weight $\lambda$. A sweep over
$\lambda$ on 1{,}764 songs confirms the prediction directly: pooled arousal $r$
falls monotonically from 0.843 to 0.810 while mean within-song $r$ rises from
0.204 to 0.368, so the same features and folds yield either a strong pooled
model or a strong dynamic one depending on nothing but the objective.

Built on this objective, MMAS fuses frozen MERT-v1-95M embeddings with ten
transcription-derived symbolic features in a windowed dual-branch TCN coupled by
cross-modal attention, trained jointly on DEAM and PMEmo. Under strict
song-grouped 5-fold cross-validation with every song out-of-fold, it reaches
pooled arousal $r = 0.801$ (DEAM) and 0.742 (PMEmo) and valence $r = 0.685$ and
0.746, at the accuracy ceiling a recent meta-analysis reports for the field
\cite{eerola2026}, and improves within-song tracking roughly threefold over a
matched linear baseline on identical features. That comparison is tested at the
song level, where each song contributes one exchangeable observation (Wilcoxon
$V = 12{,}974$, $p = 1.7 \times 10^{-21}$, Cohen's $d = 0.56$), because a
frame-level test on autocorrelated annotations would report an invalid
$p$-value. Ablations attribute independent contributions to the symbolic branch,
full MERT embeddings, and cross-modal attention, and a controlled corruption
study bounds the system's sensitivity to transcription error at the symbolic
branch's incremental contribution. Finally, a split-conformal layer replaces the
system's previously uncalibrated bands with intervals whose coverage is measured
on unseen songs ($93.7\pct$ arousal and $90.0\pct$ valence against a $95\pct$
target) and whose remaining gap, marginal rather than conditional validity, is
diagnosed rather than concealed.

All numbers are reported per dimension on unseen songs, and the trained system
is released as an open-source package so that each can be reproduced. Future
work follows directly from the limitations of Section~\ref{sec:limitations}:
group-conformal calibration treating songs as the exchangeable unit, extension
of the corpus beyond Western popular music, closing the valence gap with lyric
and cultural features, and a properly grounded tension model in the sense of
Farbood \cite{farbood2012}, validated against direct listener tension ratings
rather than derived from the two circumplex dimensions.

\section*{Reproducibility}
\addcontentsline{toc}{section}{Reproducibility}

Every number in this paper traces to a committed artifact and a script that
regenerates it.

\begin{table}[ht]
\centering
\small
\caption{Reproducibility artifacts and corresponding scripts.}
\label{tab:artifacts}

\begin{tabular}{%
  >{\raggedright\arraybackslash}p{0.28\textwidth}
  >{\raggedright\arraybackslash}p{0.38\textwidth}
  >{\raggedright\arraybackslash}p{0.26\textwidth}
}
\toprule
\textbf{Result / Analysis} &
\textbf{Artifacts} &
\textbf{Script} \\
\midrule

Table~\ref{tab:main}, Figures~\ref{fig:mainresults}--\ref{fig:persong}
(joint 5-fold OOF)
&
\url{experiments/cv_results/joint_deam_pmemo_cv.json},
\url{joint_cv_oof_predictions.csv}
&
\url{experiments/dynamic_joint_cv.py}
\\

Released-weights run (single split)
&
\url{experiments/joint_deam_pmemo_results.json}
&
\url{experiments/dynamic_joint.py}
\\

Table~\ref{tab:ablation}, Figures~\ref{fig:ablation}--\ref{fig:shape}
(ablation and $\lambda$ sweep)
&
\url{experiments/cv_results/*.json}
&
\url{experiments/dynamic_cv.py}
\\

Table~\ref{tab:significance}, Figure~\ref{fig:significance}
(significance)
&
\url{experiments/significance_results.json}
&
\url{experiments/significance_test.py}
\\

Table~\ref{tab:corruption} (corruption study)
&
\url{experiments/branch_b_corruption.json}
&
\url{experiments/branch_b_corruption.py}
\\

Table~\ref{tab:conformal} (conformal prediction)
&
\url{experiments/conformal_results.json}
&
\url{experiments/conformal_prediction.py}
\\

MERT layer probe (Figure~\ref{fig:probe})
&
\url{experiments/mert_layer_probe.json}
&
\url{experiments/probe_mert_layers.py}
\\

Feature extraction
&
MERT shards; symbolic CSVs
&
\url{experiments/extract_mert_full.py} and companions
\\

Released weights
&
\url{src/mmas/mmas_tool_bundle.pt}
&
\url{build_tool_bundle.py}
\\

All figures (1--12)
&
The artifacts listed above
&
\url{experiments/make_paper_figures.py}
\\

\bottomrule
\end{tabular}
\end{table}

Random seed 42 throughout; song-grouped splits (\texttt{GroupShuffleSplit} /
\texttt{GroupKFold}) with dataset-prefixed song IDs. All figures are regenerated
from committed result artifacts by a single script,
\texttt{experiments/make\_paper\_figures.py}; no value in any figure is entered
by hand, and none is hand-drawn except the schematic of
Figure~\ref{fig:architecture}, which contains no data. The full reproducibility
package, comprising code, committed result artifacts, and trained weights, is
publicly available at \url{https://github.com/Faraz011/MMAS}.

\section*{Declarations}
\addcontentsline{toc}{section}{Declarations}

\paragraph{Funding.}
The authors received no specific grant from any funding agency in the public,
commercial, or not-for-profit sectors.

\paragraph{Competing interests.}
The authors declare that they have no competing interests.

\paragraph{Availability of data and materials.}
DEAM: \url{https://cvml.unige.ch/databases/DEAM/}.
PMEmo: \url{https://github.com/HuiZhangDB/PMEmo}.
Code and trained model weights: \url{https://github.com/Faraz011/MMAS}.

\paragraph{Acknowledgements.}
The authors thank the DEAM and PMEmo dataset creators and the developers of
MERT, Basic Pitch, librosa, and OpenSMILE.


\end{document}